  \documentclass[preprint,review]{elsarticle}

\usepackage{fixltx2e}
\usepackage{url}
\usepackage{lineno}
\usepackage[cmex10]{amsmath}
\usepackage{amssymb}
\usepackage{slashbox}
\usepackage{color}
\usepackage{subfigure}
\usepackage{algorithm}
\usepackage{algorithmic}

\usepackage{array}
\usepackage{verbatim}
\usepackage[normalem]{ulem}
\usepackage{bbm}
\usepackage{multirow}
\usepackage[flushleft]{threeparttable}
\usepackage{tikz}

\usepackage{pifont}
\usepackage{natbib}
\usepackage{geometry}
\usepackage{fleqn}
\usepackage{graphicx}
\usepackage{txfonts}

\journal{Journal of Microprocessors and Microsystems}

\begin{document}

\begin{frontmatter}



\title{A Fair Admission Control Mechanism for Efficient Utilization of Resources in On-chip Nanophotonic Crossbars}


\author[a]{Seyed H. Mirsadeghi}
\ead{s.mirsadeghi@queensu.ca}
\author[b,c]{Ahmad Khonsari}
\ead{ak@ipm.ir}
\author[d]{M. Sadegh Talebi}
\ead{mstms@kth.se}
\author[c]{and Behnam Khodabandelo}
\ead{bk@ipm.ir}


\address[a]{Department of Electrical and Computer Engineering, Queen's University, Kingston, ON, Canada}
\address[b]{Department of Electrical and Computer Engineering, University of Tehran, Tehran, Iran}
\address[c]{School of Computer Science, Institute for Research in Fundamental Sciences (IPM), Tehran, Iran}
\address[d]{School of Electrical Engineering, The Royal Institute of Technology (KTH), Stockholm, Sweden}

\begin{abstract}
Advances in CMOS-compatible photonic elements have made it plausible to exploit nanophotonic communications to overcome the limitations of traditional NoCs. Amongst various proposed  nanophotonic architectures, optical crossbars have been shown to provide high performance in terms of bandwidth and latency. In general, optical crossbars provide a vast volume of network resources that are shared among all the cores within the chip.
In this paper, we present a fair and efficient admission control mechanism for shared wavelengths and buffer space in optical crossbars. We model buffer management and wavelength assignment as a utility-based convex optimization problem, whose solution determines the admission control policy.
Thanks to efficient convex optimization techniques, we obtain the globally optimal solution of the admission control optimization problem by using simple and yet efficient iterative algorithms. We cast our solution procedure as an iterative algorithm to be implemented a central admission controller. Our experimental results corroborate the gain that can be obtained by using such an admission controller to manage the shared resources of the system. Furthermore, they confirm that the proposed admission control algorithm works well for various traffic patterns and parameters, and evinces a tractable scalability with increase in the number of cores of the crossbar.
\end{abstract}


\end{frontmatter}


\section{Introduction}
According to Moore's law \cite{Moore}, the number of transistors on a single chip doubles every two years and this trend has made it plausible to integrate transistors on a single chip in the scale of billions. Current trend for exploiting such a large number of transistors is to have multiple computation cores on the chip rather than a single powerful processor, taking advantage of parallel computing \cite{Geer}-\cite{Parkhurst}. This has led to Chip MultiProcessor (CMP) and System-on-Chip (SoC) design paradigms with an ever increasing number of cores. In such paradigms, the communication among cores is a key design factor, which is considered to be the performance bottleneck for many-core CMPs. In fact, the communication bottleneck has made a shift in chip design from a computation-centric approach to a communication-centric one \cite{ITRS,Bjerregaard-survey}.

Network-on-Chip (NoC) was proposed as a promising design alternative to satisfy on-chip communication demands \cite{dally1910route,Benini}. A NoC design uses a packet-switched network comprising a set of routing elements connected to each other by a set of links. Despite their success in improving communication performance, traditional NoCs designed based on copper wires have been facing various challenges towards a scalable solution \cite{NoClandscape}. In the billion transistors era, wires do not scale as they occupy too much of area, and are difficult to route within the chip. In addition to electromagnetic interference problems such as the wire crosstalk, wires contribute to a large portion of the chip power consumption. Moreover, high latency in transferring messages between far nodes\footnote{Throughout the rest of this paper, we use the terms `core' and `node' interchangeably.} and the lack of sufficient bandwidth add even more limitations to traditional NoCs \cite{Owens-Dally-TradNoC}.

In accordance to the abovementioned limitations, the nanophotonic communication paradigm has gained a lot of interest as an alternative to traditional copper-based NoCs. In nanophotonic communication, data is transferred by means of light over an on-chip optical medium.
Although the idea was initially proposed four decades ago \cite{Soref}, practical limitations in optical device integration at that time put its usage off until recent years. Thanks to advances in CMOS-compatible photonic elements, optical interconnects are now considered as a practical choice for CMPs communication infrastructure \cite{Woodward,Almeida,Gunn}.

Utilizing optical interconnect can remedy many of on-chip communication restrictions faced by electrical interconnects such as crosstalk, voltage isolation, wave reflection, and etc.~\cite{Miller-Rationale}. Moreover, the high propagation speed of light in optical waveguides along with wavelength division multiplexing (WDM) techniques result in very high bandwidth and low latency communications in optical interconnects. Additionally, optical interconnects have the advantage of bit-rate transparency and low loss of optical waveguides that can highly reduce power consumption compared to their copper-wire counterparts \cite{Carloni-Emerging,Shacham}.

Various architectures have been proposed to exploit nanophotonic technology for on-chip communication networks. Among them, taking advantage of WDM techniques, optical crossbars have been shown to provide high performance in terms of bandwidth and latency. In addition, compared to other architectures, optical crossbars generally provide a vast volume of network resources shared among cores. On the other hand, NoC architectures are expected to provide different service levels to support a variety of application requirements in SoC and CMP designs. Generally speaking, in a NoC where multiple applications compete to access shared resources, a mechanism is needed to realize fair allocation of resources while guaranteeing some predefined quality of service (QoS) requirements. To this end, several researchers have studied QoS provisioning or service differentiation schemes in traditional NoCs \cite{QNoC,Radulescu,PreVirtualClock}.

The aim of this paper is to present an admission control mechanism for fair and efficient allocation of shared resources in an optical crossbar. Specifically, we consider two major resources in an optical crossbar: a) the set of wavelengths, and b) the end nodes' buffer space. Accordingly, we model wavelength assignment and buffer management in an optical crossbar as a utility-based admission control problem, using the Network Utility Maximization (NUM) framework for resource allocation in data networks \cite{Chiang_Layering}. First, we model wavelength assignment and buffer management in the crossbar as a rate control scheme. Next, we formulate the admission control policy as a rate allocation optimization problem. We consider concave utilities corresponding to applications with elastic traffic, and therefore model the admission control as a convex optimization problem. Thanks to the well-established theory of convex optimization, we obtain the globally optimal solution of the problem by using simple and yet efficient iterative algorithms. We then cast the solution as an iterative admission control algorithm to be implemented by a central controller that will be in charge of running the algorithm. Based on the results of the admission control algorithm, the central controller determines which source node may send data over each wavelength.
Our simulation results confirm that the proposed admission control algorithm works well for various traffic patterns, and evinces a tractable scalability with increase in the size of the crossbar.
Moreover, the results corroborate that the use of the proposed controller to manage shared system resources could delicately balance performance and fairness.

The rest of the paper is organized as follows. In Section \ref{sec:Rel_work}, we review the related work. Section \ref{sec:arch} describes the underlying architecture, followed by the system model description in Section \ref{sec:sys-model}.
Section \ref{sec:prob_formulation} is devoted to cast the admission control procedure as the solution to a convex optimization problem. Section \ref{sec:Opt_Sol} investigates the optimal solution, followed by the corresponding algorithms in Section \ref{sec:algs}. Simulation results are reported in Section \ref{sec:simulation}. Finally, Section \ref{sec:conclusion} concludes the paper and outlines some future directions.

\section{Related Work}
\label{sec:Rel_work}
Various approaches have been proposed to exploit nanophotonics for on-chip communication networks. Development of efficient optical data buffers is still a challenging problem, which makes it quite hard to construct a fully optical packet-switched network.
As a result, Shacham et al.~\cite{Shacham} propose a hybrid approach, in which large data packets are transferred using an optical circuit-switched network whereas the optical network is controlled by an electronic packet-switched network.
Adi et al.~\cite{Adi} consider the architecture proposed in \cite{Shacham} and propose a predictive switching and reservation technique to reduce its path setup latency. In some other hybrid architectures such as \cite{Abdollahi2016}, local communications are carried out by an electrical network, whereas long distance communications utilize optical links.
Ahmed et al.~\cite{Ahmed2015} exploit non-blocking photonic switches as well as light-weight electronic routers to decrease the latency and power overheads of hybrid architectures.
Garc\'{i}a-Guirado et al. \cite{garcia2014managing} propose a set of policies to manage hybrid networks consisting of ring-based photonic and electrical mesh sub-networks. The proposed policies use different criteria, such as message size, distance, and photonic ring availability, to decide which sub-network to be used for each message.

ATAC \cite{atac} is another hybrid architecture in which a baseline electrical 2D mesh is used for close-range point-to-point communications, whereas a ring-like optical network is used for long-distance and collective communications. The optical interconnect functions in a similar way as a broadcast bus, and contention is resolved by assigning unique wavelengths to senders. SUOR, proposed by Wu et al.~\cite{suor}, uses a circuit-switched ring-based optical NoC with a control sub-system responsible for arbitration and flow control. The control sub-system sets up the path from source to destination based on the requests received from nodes. SOUR also takes advantage of channel segmentation by dividing one waveguide into multiple non-overlapping sections that can support multiple transactions simultaneously.

Briere et al.~\cite{Briere} use 4-port optical switches to build a photonic routing structure called $\lambda$\emph{-router}, which provides contention-free communication among cores through wavelength routing techniques. Each pair of cores communicate through fixed and predefined wavelengths routed passively by the $\lambda$-router. \emph{CoNoC} is the architecture proposed by Koohi et al.~\cite{Koohi2012CoNoC}, where all-optical switches are used to implement contention-free wavelength-based passive routing of optical streams. Contention-free communication is carried out by assigning each node with a unique wavelength for data reception. This way, contention is confined to the end-points and is resolved by an electrical arbitration scheme. A scalable wavelength-routed optical NoC based on the Spidergon topology is presented in \cite{koohiSpidergon}. It uses per-receiver wavelengths in the data network to prevent network contention, and adopts per-sender wavelengths in the control network to avoid end-point contention. Werner et al.~\cite{werner2015} also propose an all-optical design exploiting a mesh-like topology.

The architecture proposed by K\i rman et al.~\cite{Kirman} is an instance of an optical crossbar where several Single Write, Multiple Read (SWMR) busses provide full connection among nodes. Each SWMR bus is dedicated exclusively to one node for sending data to others, whereas all other nodes can read data from all busses. A comparison study of worst-case optical losses for various crossbar implementations is presented in \cite{CPE:CPE3336}.
Pan et al.~\cite{Pan-Firefly} introduce \emph{Firefly}, which is a hybrid crossbar with nodes partitioned into clusters. While intra-cluster communication is done using smaller electrical crossbars, inter-cluster communication is realized by a SWMR optical crossbar. Unlike the architecture proposed in \cite{Kirman}, optical packets are not broadcast to all nodes, but rather, the intended receiver is selected by auxiliary \emph{reservation channels} prior to each communication.

\emph{Corona} is an all-optical crossbar topology, proposed by Vantrease et al.~\cite{Vantrease-Corona}, which implements a crossbar consisting of Multiple Write, Single Read (MWSR) shared busses. Each bus is dedicated exclusively to one node for receiving data, whereas all nodes can write data on all busses. Corona takes advantage of an optical token-based arbitration mechanism to resolve contention among nodes for sending data on the same bus. To address the fairness issues, a fair token slot mechanism is presented in \cite{vantrease2009light}. Fu et al.~\cite{Fu2012RCBus} employ MWSR token-ring busses of Corona to implement rows and columns of an optical 2D torus, and modified the token-based arbitration scheme so as to support virtual channel flow control.

In \cite{Ouyang-QoS}, frame-based arbitration is used to provide QoS, in terms of differentiated bandwidth allocation, for an architecture similar to Corona. The so-called \emph{FlexiShare} architecture presented in \cite{Pan-Flexishare} can be viewed as a combination of Firefly and Corona. FlexiShare makes use of Multiple Write, Multiple Read (MWMR) shared busses, where each node may write data on, or read data from, any of the shared busses. It has the advantage of lower power consumption and better channel utilization compared to other optical crossbars. Li et al.~\cite{LumiNoC} propose \emph{LumiNoC} in which the network is broken into several smaller subnets so as to avoid long waveguides. All nodes in the same subnet are connected to the same waveguide. However, communication between two subnets requires a hop through an intermediate electrical router.

Closest to our work is the work done by Pan et al., which present \emph{FeatherWeight} \cite{FeatherWeight}. FeatherWeight provides an optical arbitration scheme with QoS support in nanophotonic MWSR crossbars. Similarly to \cite{vantrease2009light}, they use token streams to grant access to source nodes for sending data to each home (destination) node. In addition, each source node is assigned a ``quota" that designates the maximum number of tokens a node can grab in each time slot. To improve resource utilization and enforce QoS, the quotas are dynamically changed with respect to the request patterns. A QoS controller residing at each home node is used to update the quotas. At each time slot, every node gives feedback to the controller indicating the amount of tokens it has consumed in the previous time slot. Having gathered the feedback, the controller first updates the quotas and then propagates the updated values to the nodes. Access to data channels is granted to nodes with respect to the new quotas in the following time slot. The proposed QoS mechanism provides both fairness and differentiated service among the nodes.

Similarly to FeatherWeight, we propose a time-slotted mechanism to enable fairness and differentiated service among the nodes of a nanophotonic crossbar. However, FeatherWeight is based on an MWSR architecture, whereas we target MWMR photonic crossbars. In FeatherWeight, each MWSR channel is arbitrated separately, and fairness is enforced on each MWSR channel independently of other MWSR channels. We present an admission control scheme that takes into account all available channels in the optical crossbar as a whole, and provides fairness and QoS globally among all nodes and all data channels. In contrast to FeatherWeight that only provides max-min fairness, our channel allocation scheme could cover a larger class of fairness metrics. Finally, we use a centralized controller to carry out resource allocation among the nodes. We use a different QoS algorithm that models the problem as an optimization problem whose solution is the optimal resource allocation in each time slot.

\section{Nanophotonic Architecture}
\label{sec:arch}
\begin{figure}
	\begin{center}
		\includegraphics[angle=0, scale=0.40]{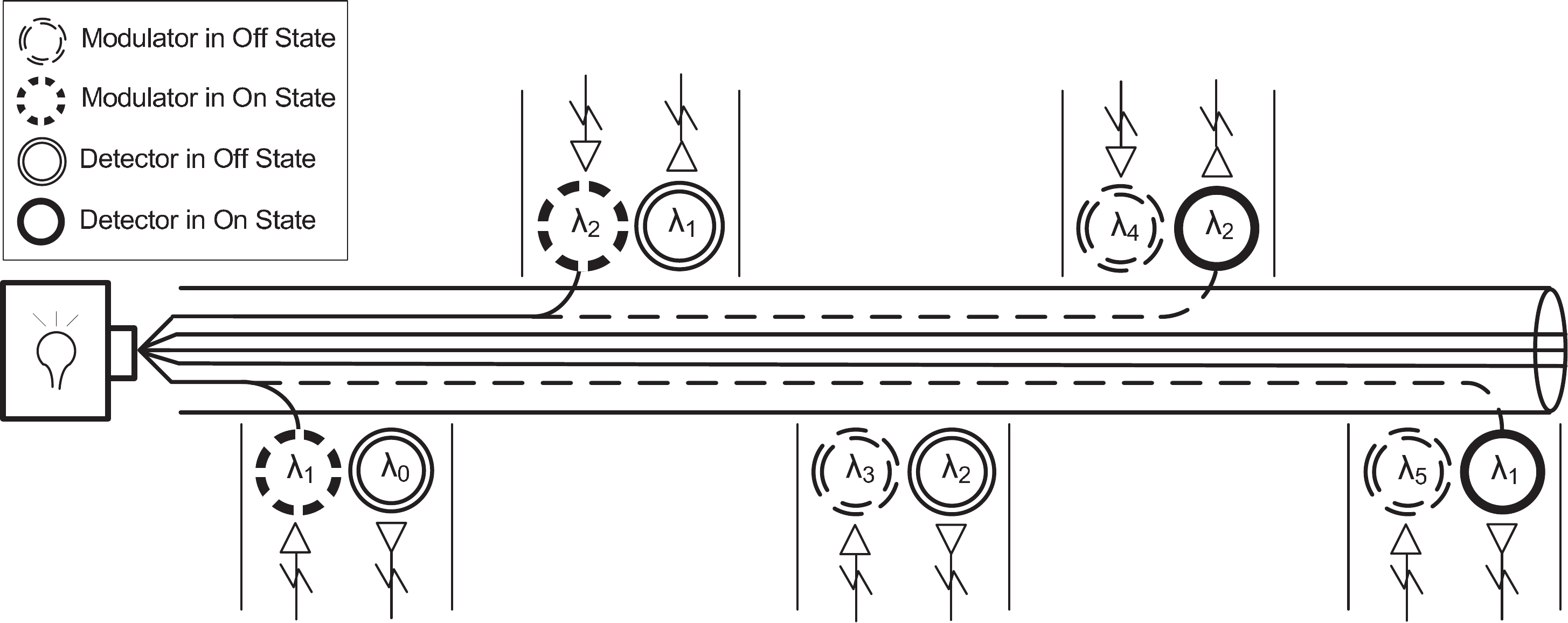}
	\end{center}
	\caption{Basic configuration of nanophotonic elements for optical data transmission}
	\label{fig:optComDiagram}
\end{figure}

Three basic blocks are needed to realize a nanophotonic interconnect: \emph{silicon waveguides, microring resonators}, and \emph{laser sources}. Waveguides are the medium over which the optical signals are transferred at the speed of light. Using DWDM\footnote{Dynamic Wavelength Division Multiplexing} techniques, multiple wavelengths can travel in a single waveguide simultaneously without interfering with each other. Microring resonators are the dominant state-of-the-art elements used in optical NoCs for modulation and/or detection of particular wavelengths. A light generation source is also required to provide the beam of light over which data is modulated. Due to difficulties in integrating a silicon-based laser onto a chip, off-chip laser sources are preferred. The light generated off-chip is coupled onto the chip by means of optical fibers. Figure \ref{fig:optComDiagram} shows a basic configuration of nanophotonic elements for optical on-chip communications.

\begin{figure}[t]
	\centering
	\subfigure[Single-Round Channels]{\includegraphics[scale=0.38]{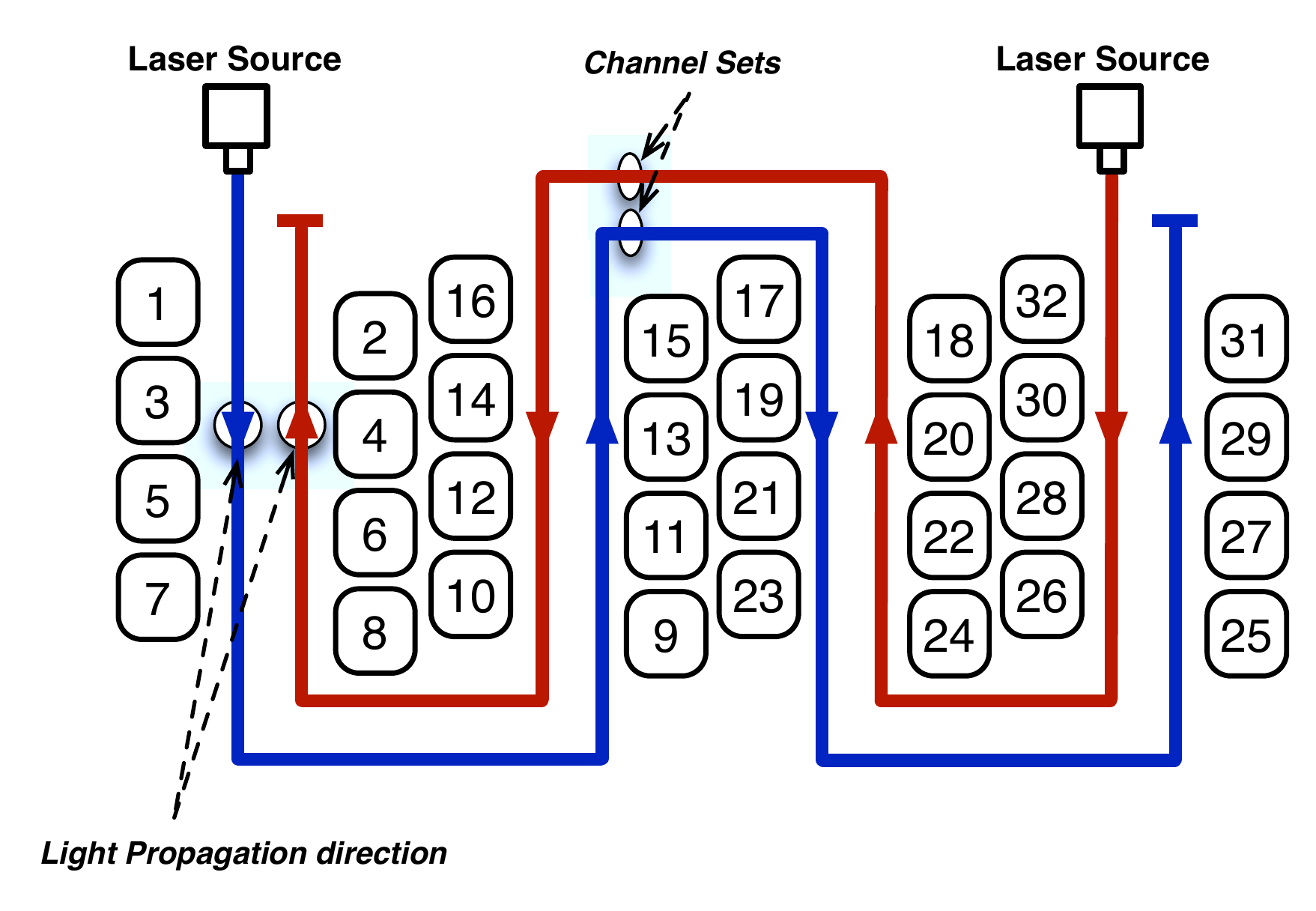}
		\label{fig:layout1}
	}
	\subfigure[Two-round Channels]{\includegraphics[scale=0.38]{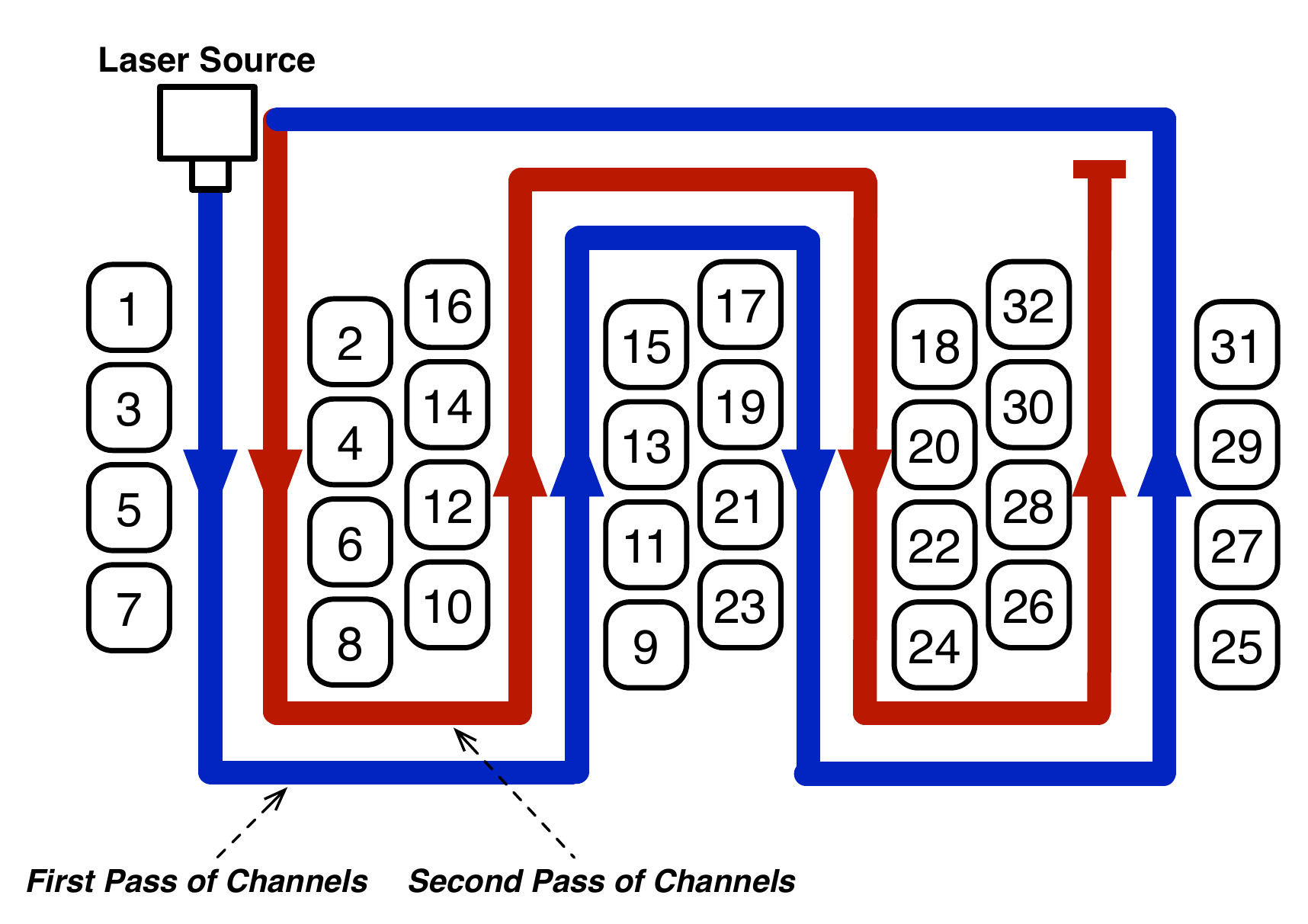}
		\label{fig:layout2}
	}
	\caption{Layout of waveguides}
	\label{fig:layout-Diagram}
\end{figure}

The architecture we consider in this work is similar to the one proposed in \cite{Pan-Flexishare} referred to as Flexishare. As mentioned in Section \ref{sec:Rel_work}, it uses Multiple Write, Multiple Read (MWMR) shared busses to implement a crossbar among the nodes, and each node may write or read data to/from any of the shared busses. In such an architecture, the whole network can be considered as a pool of communication channels shared among all nodes that can be dynamically configured to support data transmission between multiple source-destination pairs. From nanophotonic point of view, channels are realized through several waveguides deployed on the chip. We consider the same waveguide layout presented in \cite{Joshi-PhotonicClos}, where each waveguide starts its path from the upper left corner of the chip and visits all nodes along a serpentine path that terminates at the upper right corner. Figure \ref{fig:layout1} depicts a schematic diagram of such a layout for a 4$\times$8 NoC.

Terminating waveguides at the upper right corner leads to a \emph{single-round} implementation of data channels as each waveguide passes all nodes exactly once. To provide full connectivity, each node needs to modulate light in opposite directions on the waveguides. Therefore, the channels should be divided into two sets, where in one set light is injected from the upper left corner of Figure \ref{fig:layout1} propagating toward the upper right corner, and vice versa for the other set. In Figure \ref{fig:layout1}, the two sets have been shown in blue and red, respectively. Depending on the relative position of the source and destination nodes (which dictates the required direction of light transmission), channels from the first or the second set will be used.

In Figure \ref{fig:layout1}, all nodes have been indexed from 1 to 32. An arbitrary source node with index $i$ can use the set of channels shown in blue for sending data to destination node $j$ only if $i < j$. Similarly, red channels can be used only when $i>j$. Consequently, though a single-round layout of waveguides benefits from shorter length and lower power dissipation, it imposes some limitation in terms of channel sharing because the channels in one set cannot be utilized for data transmission in the opposite direction. To clarify this, consider the scenario in which nodes $1,2,\dots,31$ wish to send data to nodes $2,3,\dots,32$, respectively. Since each source node has a lower index than its corresponding destination node, all the senders can only utilize the set of blue channels for data transmission. This results in a high contention over the channels in one set, as well as a low utilization of the channels in the other set.

In order to mitigate this problem, which limits our ultimate goal of total sharing of channels, we use another implementation of optical data channels called \emph{two-round} channels \cite{Pan-Flexishare}. The waveguide layout is the same as the one shown in Figure \ref{fig:layout1}, but each waveguide continues its path back to the originating point at the upper left corner rather than being terminated at the upper right corner. Figure \ref{fig:layout2} portrays an instance of such a layout. In this way, each waveguide will pass each node twice along its path, and the light will only be injected in one direction. For all nodes, the first pass of each waveguide (shown in blue in Figure \ref{fig:layout2}) may only be used for light modulation, whereas detection takes place only on the second pass (shown in red). Thus, a node can utilize any of the available channels across all waveguides for data transmission to any other node irrespective of the relative position of the sender and receiver. With the same number of waveguides, a two-pass layout will potentially provide each node with more channels for data transmission compared to a single-pass layout. This is shown by using thicker lines in Figure  \ref{fig:layout2} to illustrate the waveguides.

With the above-mentioned design, each node can potentially make use of all the channels for data reception and/or transmission. This brings a higher flexibility in terms of channel sharing among the nodes, which in turn provides better utilization of channels. A major challenge in such an architecture will be the efficient management and allocation of channels. We will deeply explore this issue in the following sections.

\section{System Model}
\label{sec:sys-model}
We consider the architecture described in Section \ref{sec:arch} with $N$ nodes indexed by $n\in\{1,\dots,N\}$, and $W$ waveguides deployed in a two-round fashion. We assume 64 wavelengths are multiplexed on each waveguide providing $C = 64 \times W$ channels in the system. Letting $R$ denote the data rate of an individual wavelength, the transmission rates will be multiples of $R$.
In order to transmit data, each sending node should be allocated at least one of the $C$ wavelengths. Allocating more wavelengths to a node provides it with a higher rate for data transmission. In addition, each wavelength can be allocated to at most one node at any moment.

We partition time into multiple slots of length $\delta$. Let $\mathcal I^t$ denote the set of all nodes that have data for transmission at time slot $t$. We define $\mathcal I_k^t\subset\mathcal I^t$ as the set of nodes wishing to send data to destination node $k$. Equivalently, node $k$ is the destination node for the set $\mathcal I_k^t$. Moreover, we assume that each node will have some
specific buffer space for receiving data from other nodes. For any node $k$, the length of this buffer depends on both the cumulative rate at which other nodes send data to node $k$, and the drain rate of node $k$. Drain rate of a node denotes the rate at which a node can process received data. We assume that at time slot $t$, node $k$ can process incoming packets with rate $r_k^t$, has $M_k^t$ free buffer space, and receives data from node $n\in\mathcal I^t_k$ at rate $x_{nk}^t$. It is worth mentioning that $x_{nk}^t$ is proportional to the number of wavelengths (channels) assigned to each node for data modulation.

We represent by $\boldsymbol x_n^t=(x_{nk}^t, k=1,\dots,N)$ the rate allocation vector or simply rate vector for (sender) node $n$ at time $t$.
We also denote the overall rate allocation at time slot $t$ by the vector $\mathbf X^t=[\boldsymbol x_1^t ,\dots, \boldsymbol x_N^t]$ made by concatenating rate vectors of all (sender) nodes.
The rate allocation policy at time slot $t$ is feasible if and only if the corresponding rate allocation vector $\mathbf X^t\in\mathbb R^{N^2}$ satisfies:
\begin{align}
\label{eq:channel_aloc_feas0_leq}
&x_{nk}^t\ge 0;\quad n,k=1,\dots,N,\\
\label{eq:channel_aloc_feas1_leq}
&\sum_{n\in\mathcal I_k^t} x_{nk}^t\leq r_k^t + \frac{M_k^t}{\delta}; \quad k=1,\dots,N,\\
\label{eq:channel_aloc_feas2_leq}
&\sum_{n\in\mathcal I^t} \sum_{k=1, k\neq n}^{N} x_{nk}^t \leq CR.
\end{align}
The first condition requires that all allocated rates be nonnegative.
The term $\frac{M_k^t}{\delta}$ in the r.h.s.~of (\ref{eq:channel_aloc_feas1_leq}) represents the rate at which a receiving node can buffer the incoming packets, i.e., the receiver buffering rate. Thus, the buffering rate of a receiver is the rate at which free buffer space is available to its transmitting node(s).
The second equation accounts for feasibility of rate allocation in terms of flow control, and provides an upper bound for the cumulative rate at which nodes can receive data without buffer overflows. This upper bound depends on both the drain and buffering rates. Thus, (\ref{eq:channel_aloc_feas1_leq}) states that for each receiving node, the cumulative rate of incoming data from all other senders should be no greater than the sum of drain and buffering rates\footnote{Indeed, we assume that each node knows its available buffer space and drain rate a priori. In a practical setting, however, each node might need to predict them to be within some specified error.}.
The third equation implicitly accounts for feasibility in terms of wavelength availability. It states that the sum of the rates assigned to all nodes for data transmission cannot be greater than the total rate provided by the pool of wavelengths.
As long as this condition holds, there will be enough channels available to each node with the assigned rate, provided that each channel is allocated to one sender at most.
For the sake of convenience in our derivations in later sections, we introduce \textit{sender-receiver traffic matrix} or simply \textit{traffic matrix} $\mathbf A^t=[a_{nk}^t]_{N\times N}$ with $a_{nk}^t=1$ if $n\in\mathcal I_k^t$ and $a_{nk}^t=0$ otherwise.
Indeed, $\mathbf A$ represents the traffic pattern of the whole system in terms of the communications between all source-destination pairs.

\section{Admission Control Optimization Problem}
\label{sec:prob_formulation}
In this section, we use the rate allocation model discussed in Section \ref{sec:sys-model} to formulate the admission control problem as a rate allocation optimization problem that finds the best rate allocation among all feasible rate vectors in view of (\ref{eq:channel_aloc_feas0_leq})-(\ref{eq:channel_aloc_feas2_leq}).
We seek to derive a rate allocation policy which leads to a good channel utilization by using as many available channels (wavelengths) as possible, as well as a fair assignment of the resources to nodes. To this end, we cast the admission control policy as the solution to a utility-based optimization problem that strives to find a rate vector with the highest satisfaction among the set of all feasible rate vectors.
In order to quantify the satisfaction level of nodes, we use the notion of utility function. More specifically, let $U_{nk}(.)$ denote the utility function assigned to each logical connection $(n,k)$. This implies that if at time slot $t$, node $n$ sends data to node $k$ at rate $x_{nk}^t$, it will attain a satisfaction level quantified by $U_{nk}(x_{nk}^t)$. We assume that the utility function $U_{nk}(.)$ satisfies the following conditions \cite{Mo2000utility}:

\begin{itemize}
	\item[\textbf{C1:}] The function $U_{nk}(.)$ is continuous, increasing, and twice differentiable over $(0,\infty)$.
	\item[\textbf{C2:}] The function $U_{nk}(.)$ is strictly concave with bounded curvature.
\end{itemize}

It is worth mentioning that the above conditions are not restrictive as our focus is on applications that admit elastic traffic demand. In other words, traffic characteristics of such applications can be efficiently captured by a utility function satisfying conditions \textbf{C1} and \textbf{C2}.
A well-known class of utility functions that satisfies conditions \textbf{C1} and \textbf{C2} is the class of $\alpha$-fair utility functions introduced in \cite{Mo2000utility}, where every utility function is defined based on a fairness parameter $\alpha>0$ as
\begin{align}
\label{eq:alphafair_utility}
U(x,\alpha)=\left \{ \begin{array}{ll}
w\frac{x^{1-\alpha}}{1-\alpha} & \alpha\neq 1\\
w\log x & \alpha=1\\
\end{array} \right.,
\end{align}
where $w$ is a positive weight.
The choice of $\alpha$ determines the tradeoff between the throughput and fairness. For larger values of $\alpha$, the system sacrifices the throughput to allocate resources in a \emph{fairer} way. As $\alpha\to\infty$, we approach the max-min fair allocation, which yields the best fairness at expense of the worst throughput.

Taking into consideration all receiving nodes to which node $n$ may have data to send, we define the utility of node $n$ as the sum of the utility functions for its logical connections as
\begin{align}
U_n(\boldsymbol x^t_n) = \sum_{k=1}^N a_{nk}^tU_{nk}(x^t_{nk}).\nonumber
\end{align}
We then define the utility of the whole crossbar as
\begin{align}
\label{eq:total_utility_new}
U(\mathbf X^t)=\sum_{n=1}^N U_n(\boldsymbol x_n^t)
=\sum_{n=1}^N \sum_{k=1}^N a_{nk}^tU_{nk}(x_{nk}^t).
\end{align}

In the rest of this paper, we focus on the time slot $t$, and hence omit the superscript $t$ hereafter. Following the Network Utility Maximization (NUM) framework (see, e.g., \cite{Chiang_Layering}), our objective is to assign available wavelengths to senders so as to maximize the total utility, defined in (\ref{eq:total_utility_new}), over all feasible rate vectors:
\begin{align}
\label{eq:opt_prb}
\max_{\mathbf X\ge 0}& \;\; \sum_n\sum_k a_{nk}U_{nk}\left(x_{nk}\right)\\
\label{eq:constraint1}
\mathrm{subject~to:}&\;\;\sum_{n=1}^{N} a_{nk}x_{nk}\leq r_k + \frac{M_k}{\delta}; \quad k=1, \dots, N,\\
\label{eq:constraint2}
&\sum_{n=1}^N \sum_{k=1}^{N} a_{nk}x_{nk} \leq CR.
\end{align}

The optimization variable in (\ref{eq:opt_prb}) is the vector $\mathbf X$.
We denote the optimal rate vector for the above problem by $\mathbf X^\star=[\boldsymbol x_1^\star ,\dots, \boldsymbol x_N^\star]$, where $\boldsymbol x_n^\star=(x_{nk}^\star,k=1,\dots,N),$ for $n=1,\dots,N$. The optimal rate vector determines the optimal admission control policy.
Due to conditions \textbf{C1}-\textbf{C2}, the objective function of problem (\ref{eq:opt_prb})-(\ref{eq:constraint2}) is strictly concave as it is a nonnegative sum of strictly concave functions. Moreover, its constraints are affine functions. Thus, problem (\ref{eq:opt_prb})-(\ref{eq:constraint2}) is strictly convex \cite{NLP_Bertsekas}. Furthermore, the feasible region, i.e., the polyhedron defined by constraints (\ref{eq:constraint1})-(\ref{eq:constraint2}), is connected and bounded, and hence is a compact set. Thus, at least one optimal solution exists. Finally, the maximizer is unique due to the strict convexity of the problem \cite{NLP_Bertsekas}.

\section{Optimal Solution}
\label{sec:Opt_Sol}
This section is devoted to solving problem (\ref{eq:opt_prb})-(\ref{eq:constraint2}). The standard approach to solve such a constrained convex problem is to solve the (unconstrained) dual of problem . Duality theory guarantees that under mild conditions, which hold for our problem, solving the dual yields the optimal solution to the primal problem. To do to, in what follows we first establish its Lagrangian and derive the dual function. We then formulate the dual problem and solve it using iterative methods.

\subsection{Primal Optimality Analysis}
We start by writing the Lagrangian for problem (\ref{eq:opt_prb})-(\ref{eq:constraint2}) \cite{BV_cvx}:
\begin{align}
	\label{eq:Lagrangian}
	L(\mathbf{X},\boldsymbol\lambda)=\sum_n\sum_k a_{nk}U_{nk}(x_{nk})
	-\lambda_0\left(\sum_{n=1}^N \sum_{k=1}^{N} a_{nk}x_{nk}-CR\right)
	-\sum_{k=1}^N\lambda_k\left(\sum_{n=1}^{N} a_{nk}x_{nk}-r_k-\frac{M_k}{\delta}\right).
\end{align}
Here, $\boldsymbol\lambda=(\lambda_k,k=0,\dots,N)$ is the vector of positive Lagrange multipliers associated with constraints (\ref{eq:constraint1}) and (\ref{eq:constraint2}). According to convex optimization theory, the primal-optimal vector $\mathbf{X^\star}$ and dual-optimal vector $\boldsymbol\lambda^\star$ must satisfy Karush-Kuhn-Tucker (KKT) conditions \cite{BV_cvx}:
\begin{align}
&\nabla_{\mathbf X} L(\mathbf{X}^\star,\boldsymbol\lambda^\star)=\mathbf 0,\;\;\; \boldsymbol\lambda^\star\geq \mathbf 0,\nonumber\\
\label{eq:primal_feasibility2}
&\sum_{n=1}^{N} a_{nk}x^\star_{nk}\leq r_k + \frac{M_k}{\delta}; \quad k=1, \dots, N,\\
\label{eq:primal_feasibility22}
&\sum_{n=1}^N \sum_{k=1}^{N} a_{nk}x^\star_{nk} \leq CR,\\
\label{eq:KKT_constraint_total}
&\lambda^\star_0\left(\sum_{n=1}^N \sum_{k=1}^{N} a_{nk}x^\star_{nk} - CR\right)=0,\\
\label{eq:KKT_constraint_k}
&\lambda^\star_k\left(\sum_{n=1}^{N} a_{nk}x^\star_{nk} - r_k - \frac{M_k}{\delta}\right)=0; \; k=1, \dots, N.
\end{align}

Expanding the first KKT condition, we have for any $k$ and $n$:
\begin{align}
\label{eq:staitionary_cond}
\frac{\partial L}{\partial x^\star_{nk}}=a_{nk}U^\prime_{nk}(x^\star_{nk})-a_{nk}\lambda_k^\star-a_{nk}\lambda_0^\star=0,
\end{align}
which further gives
\begin{align}
\label{eq:x_opt}
x_{nk}^\star=\bigg[a_{nk}U_{nk}^{\prime-1}\left(\lambda^\star_0+\lambda^\star_k\right)\bigg]^+,
\end{align}
where $[z]^+=\max\{z,0\}$. Having computed $\mathbf {X^\star}$, we define the dual function $D(\cdot)$ as \cite{BV_cvx}:
\begin{align}
\label{eq:dualFunc}
D(\boldsymbol\lambda):=\max_{\mathbf X} L(\mathbf X,\boldsymbol\lambda)=L(\mathbf X^\star,\boldsymbol\lambda).
\end{align}
We note that Lagrange multipliers are referred to as dual variables since the dual function $D(\cdot)$ is a function of Lagrange multipliers.
The dual-optimal vector $\boldsymbol \lambda^\star$ could be found by solving the dual problem of problem (\ref{eq:opt_prb})-(\ref{eq:constraint2}),
given by \cite{BV_cvx}:
\begin{align}
\label{eq:dual_problem}
\min_{\boldsymbol\lambda\geq \mathbf 0} D(\boldsymbol\lambda).
\end{align}
Next, we solve dual problem (\ref{eq:dual_problem}) using iterative methods. Due to strict convexity of the primal problem (\ref{eq:opt_prb})-(\ref{eq:constraint2}), $D(\cdot)$ is continuously differentiable with partial derivatives given by Danskin's Theorem \cite{NLP_Bertsekas}:
\begin{align}
\label{eq:partialD_1}
\frac{\partial D}{\partial \lambda_k}=& r_k + \frac{M_k}{\delta}-\sum_{n=1}^{N} a_{nk}x_{nk}; \quad k=1, \dots, N,\\
\label{eq:partialD_2}
\frac{\partial D}{\partial\lambda_0}=&CR-\sum_{n=1}^N \sum_{k=1}^{N} a_{nk}x_{nk}.
\end{align}
Thus, we take the advantage of using simple iterative methods such as gradient projection algorithm and its variants. In order to achieve a fast yet simple iterative algorithm, we use \emph{diagonally scaled gradient projection algorithm} \cite{NLP_Bertsekas} to solve problem (\ref{eq:dual_problem}). This algorithm can be seen as an approximate to the Newton's method and hence we expect that it will converge faster than the gradient method while exhibiting tractable scalability features.

Using this algorithm, the dual variable update equation at $m$-th iteration is given by
\begin{align}
\label{eq:dual-update}
\boldsymbol\lambda^{(m+1)}=\bigg[\boldsymbol\lambda^{(m)}-\gamma^{(m)}\mathbf B^{(m)}\nabla D\left(\boldsymbol\lambda^{(m)}\right)\bigg]^+,
\end{align}
where $\gamma^{(m)}$ is a step size and $\mathbf B^{(m)}$ is a diagonal matrix whose main diagonal elements are the inverse of second partial derivatives of the dual function:
\begin{align}
\label{eq:matrixB}
\mathbf B=\textrm{\textbf{diag}}\left(\left(\frac{\partial^2 D}{\partial\lambda_k^2}\right)^{-1}, \; k=0,\dots,N\right).
\end{align}
Using (\ref{eq:partialD_1})-(\ref{eq:partialD_2}), the required second derivatives of the dual function stated in (\ref{eq:dualFunc}) are obtained below:
\begin{align}
\label{eq:matrixB_1}
\frac{\partial^2 D}{\partial\lambda_k^2}&=-\sum_{n=1}^{N} a_{nk}\frac{\partial x_{nk}}{\partial\lambda_k}
= -\sum_{n=1}^{N} a_{nk}\frac{\partial}{\partial\lambda_k}U_{nk}^{\prime-1}\left(\lambda_0+\lambda_k\right)
= -\sum_{n=1}^{N} \frac{a_{nk}}{U_{nk}^{\prime\prime}\left(x_{nk}\right)}, \quad k=1,\dots,N, \\
\label{eq:matrixB_2}
\frac{\partial^2 D}{\partial\lambda_0^2}&=-\sum_{n=1}^{N}\sum_{k=1}^N a_{nk}\frac{\partial x_{nk}}{\partial\lambda_0}
= -\sum_{n=1}^{N}\sum_{k=1}^N a_{nk}\frac{\partial}{\partial\lambda_k}U_{nk}^{\prime-1}\left(\lambda_0+\lambda_k\right)
= -\sum_{n=1}^{N}\sum_{k=1}^N \frac{a_{nk}}{U_{nk}^{\prime\prime}\left(x_{nk}\right)}.
\end{align}

Step size $\gamma^{(m)}$ must be chosen so as to guarantee the convergence of the iterative algorithm while yielding fast convergence. One of the best choices of step size is the \emph{diminishing step size rule} that satisfies \cite{NLP_Bertsekas}:
\begin{align*}
	\gamma^{(m)} \geq 0, \; \forall m\in \mathbb N,\;\;\;
	\lim_{m\rightarrow\infty} \gamma^{(m)}=0,\;\;\;
	\sum_{m=1}^\infty \gamma^{(m)}=\infty.
\end{align*}
In this paper we use $\gamma^{(m)}=\frac{d}{\sqrt{m}}$ with some constant $d>0$ (to chosen later), which satisfies the above conditions. Substituting partial derivatives into (\ref{eq:dual-update}), update equations for dual variables $\lambda_0$ to $\lambda_N$ are given by:
\begin{align}
	\label{eq:dual-update_elementwise_1}
	\lambda_0^{(m+1)}&= \bigg[\lambda_0^{(m)}+\frac{\gamma^{(m)}}{\sum_{n=1}^{N}\sum_{k=1}^N \frac{a_{nk}}{U_{nk}^{\prime\prime}\left(x_{nk}^{(m)}\right)}}\left(CR-\sum_{n=1}^N \sum_{k=1}^{N} a_{nk}x_{nk}^{(m)}\right)\bigg]^+,\\
	\label{eq:dual-update_elementwise_2}
	\lambda_k^{(m+1)}&= \bigg[\lambda_k^{(m)}+\frac{\gamma^{(m)}}{\sum_{n=1}^{N} \frac{a_{nk}}{U_{nk}^{\prime\prime}\left(x_{nk}^{(m)}\right)}}\left(r_k + \frac{M_k}{\delta}-\sum_{n=1}^{N} a_{nk}x_{nk}^{(m)}\right)\bigg]^+, \quad k=1,\dots,N,
\end{align}
where $x^{(m)}_{nk}$ is the value for the primal variables at the $m$-th iteration of the algorithm given by
\begin{align}
\label{eq:genPrimalUpdate}
\mathbf X^{(m+1)}=\arg\max_{\mathbf X} L(\mathbf X,\boldsymbol\lambda^{(m)}).
\end{align}
Equivalently, $x^{(m)}_{nk}$ is given by (\ref{eq:x_opt}) when $\boldsymbol\lambda^{(m)}$ is used instead $\boldsymbol\lambda^\star$ as its approximation until the $m$-th iteration.

Using the iterations outlined in (\ref{eq:dual-update_elementwise_1}) and (\ref{eq:dual-update_elementwise_2}), the sequence $\{\boldsymbol\lambda^{(m)}\}$ will converge to dual-optimal variables $\boldsymbol\lambda^\star$. Because of strict convexity of primal problem (i.e., (\ref{eq:opt_prb})-(\ref{eq:constraint2})), strong duality holds and hence it is guaranteed that both dual and primal problems will have the same optimal objective \cite{BV_cvx}. Thus, solving the dual problem (\ref{eq:dual_problem}) leads to the optimal solution of the primal and guarantees that the sequence $\{\mathbf X^{(m)}\}$ obtained by (\ref{eq:genPrimalUpdate}) converges to the optimal solution of the admission control problem. We defer the algorithmic aspects of this iterative solution until the next section.

\subsection{$\boldsymbol\alpha$-Fair Utility Functions}
Now we concentrate on the case of $\alpha$-fair utility functions. First, we consider the case of $\alpha\neq 1$. Denoting by $\mathbf X^\star(\alpha)$ the optimal rate vector for utility with parameter $\alpha$, using (\ref{eq:alphafair_utility}) and (\ref{eq:x_opt}) we have
\begin{align}
\label{eq:x_opt_alpha}
x_{nk}^\star(\alpha)=a_{nk}w_{nk}\left(\lambda_k+\lambda_0\right)^{-\frac{1}{\alpha}}.
\end{align}

Furthermore, calculation of partial derivatives for $\alpha$-fair utility functions produces update equations for this class of utility functions given by (\ref{eq:dual-update_elementwise_1}) and (\ref{eq:dual-update_elementwise_2}) with $U''(x_{nk}^{(m)})=w_{nk}[x_{nk}^{(m)}\big]^{-(\alpha+1)}$.
It is worth noting that $1$-fair utility (\mbox{$\alpha=1$}) can be obtained by calculating the limit of $\alpha$-fair utility for \mbox{$\alpha\neq 1$} when $\alpha$ approaches 1:
\begin{align}
	\label{eq:x_opt2_prop}
	x_{nk}^\star(1)&=\lim_{\alpha\rightarrow 1} x_{nk}^\star(\alpha)=\frac{a_{nk}w_{nk}}{\lambda_k+\lambda_0}.
\end{align}
Accordingly, dual variable updates for $\alpha=1$ can be obtained by asserting $\alpha=1$ into (\ref{eq:dual-update_elementwise_1}) and (\ref{eq:dual-update_elementwise_2}).
Thus, when describing the admission control algorithm in later sections, equations (\ref{eq:x_opt_alpha})-(\ref{eq:dual-update_elementwise_2}) are valid for all $\alpha >0$.


\section{Admission Control Algorithm}
\label{sec:algs}
Although the above-mentioned optimization procedure for the admission control might look complicated, it could practically be implemented using the algorithms that will be described in this section. To gain more insights into the design of algorithms, we would only concentrate on the case of $\alpha$-fair utility functions. We note, however, that one can simply use (\ref{eq:x_opt}), (\ref{eq:dual-update_elementwise_1}), and (\ref{eq:dual-update_elementwise_2}) when using any other utility functions satisfying conditions \textbf{C1} and \textbf{C2}. In the sequel, we first present an algorithm for the iterative calculation of primal-optimal and dual-optimal vectors. Second, based on the obtained optimal values, we devise our final wavelength assignment algorithm by means of a centralized admission controller.

\subsection{Iterative Solution to Admission Control Problem}
\label{sec:alg1}
Algorithm 1 lists the required steps to solve our admission control optimization problem (\ref{eq:opt_prb})-(\ref{eq:constraint2}) by iteratively solving its dual problem (\ref{eq:dual_problem}). This algorithm includes the iterations given by (\ref{eq:x_opt_alpha}), (\ref{eq:dual-update_elementwise_1}), and (\ref{eq:dual-update_elementwise_2}), and begins by choosing an initial feasible value for $\mathbf X$ and $\boldsymbol\lambda$. As long as the specified stopping criterion (defined below) is not met, at each iteration, the dual variable vector $\boldsymbol\lambda^{(m)}$ is updated first. Based on the updated dual variables, the algorithm calculates $\mathbf X^{(m)}$. Eventually the stopping criterion is met, and then $\mathbf X^{(m)}$ and $\boldsymbol\lambda^{(m)}$ are reported as the approximate values of $\mathbf X^{\star}$ and $\boldsymbol\lambda^{\star}$, respectively.

One interesting performance metric is the convergence behavior of the proposed algorithm. Of particular concern is the number of iterations required for the algorithm to converge. Gradient-like algorithms do not converge after a finite number of iterations \cite{NLP_Bertsekas}, and hence it is necessary to determine a stopping criterion for the algorithm based on some predefined accuracy that can approximate the distance from the globally optimal point.

\begin{algorithm}[h]
\small
   \caption{Iterative Solution}
   \label{alg:iterative_solution}
\begin{algorithmic}
   \STATE {\bf Initialization:}
   \STATE Choose feasible starting points $\mathbf{X}^{(0)}$ and $\boldsymbol\lambda^{(0)}$. Set $m=1$.
   \WHILE{$\max_{n,k}|x^{(m+1)}_{nk}-x^{(m)}_{nk}|>\epsilon R$}

    \STATE Set step size $\gamma^{(m)}=\frac{d}{\sqrt{m}}$.
    \STATE Update dual variables using (\ref{eq:dual-update_elementwise_1}) and (\ref{eq:dual-update_elementwise_2}).
    \STATE Update primal variables for all $k$ and $n$: \;\;  $x_{nk}^{(m+1)}=\bigg[a_{nk}w_{nk}\left(\lambda_k^{(m+1)}+\lambda_0^{(m+1)}\right)^{-\frac{1}{\alpha}}\bigg]^+$.
    \STATE Increment $m$: $m=m+1$.
   \ENDWHILE
\end{algorithmic}
\normalsize
\end{algorithm}

The stopping criterion for the iterative algorithm is chosen as follows. Given $\epsilon\in(0,1)$, we terminate the algorithm when the largest change in source rates is less than $\epsilon R$, i.e.,
\begin{align}
\label{eq:stop_crit}
\max_{n,k}|x_{nk}^{(m+1)} - x_{nk}^{(m)}| \leq \epsilon R.
\end{align}
We then define $I_\epsilon\triangleq \min\left\{m:\max_{n,k}|x_{nk}^{(m+1)} - x_{nk}^{(m)}| \leq \epsilon R\right\}$ to denote the minimum number of iterations required for the algorithm to meet the stopping criterion mentioned above.
We note that there is a tradeoff in choosing $\epsilon$. A lower value of $\epsilon$ guarantees that the final result will be closer to the globally optimal point, however, at the expense of larger $I_\epsilon$, i.e., more iterations.

\subsubsection{Numerical Experiments}
In order to quantify $I_\epsilon$, we need to gain some insight into how the problem parameters would affect $I_\epsilon$. To this end, we have carried out several simulation experiments through implementing Algorithm \ref{alg:iterative_solution} in MATLAB. In particular, we are interested in the influence of the following parameters on $I_\epsilon$:
\begin{enumerate}
	\item Number of nodes in the crossbar ($N$)
	\item Density and pattern of traffic matrix ($\mathbf A$)
	\item Step size ($\gamma$)
\end{enumerate}

In all experiments, we consider a MWMR crossbar consisting of 32 waveguides each carrying 64 wavelengths. Considering 10 Gbps transmission rate for each wavelength, the total capacity of the crossbar is $C=20.48$ Tbps. Moreover, for all nodes, the values for the parameters $r_k$ and $w_{nk}$ are drawn independently and uniformly at random from the intervals $[0, C]$ and $[0, 1]$, respectively. Each node is assumed to have free buffer space for storing $g$ packets, where $g$ is chosen from the set $\{1,2,\dots,20\}$ uniformly at random.
Finally, we set $\delta=5.4$ ns and $\epsilon=10^{-11}$.

We consider crossbars with $N=64, 128, \textrm{and } 256$ number of nodes, and use a diminishing step size $\gamma=\frac{d}{\sqrt m}$ with $d=3,5,\textrm{and } 7$. Since both pattern and density of the traffic matrix $\mathbf{A}$ may influence $I_\epsilon$, we use randomly generated traffic matrices with different densities, ranging from sparse to very dense. The matrix density represents the ratio of non-zero elements (equivalently 1s) to all $N^2$ elements of $\mathbf {A}$. Thus, for example 2\% density means that $0.02 N^2$ elements of $\mathbf {A}$ are 1. To take into account the randomness of problem parameters, for each choice of $(N,\mathbf A,\gamma)$, we report the empirical average of $I_\epsilon$, denoted by $\overline I_\epsilon$, measured over $100$ experiments.
\begin{figure}[htp]
	\centering
	\subfigure[Crossbar with $N=64$ nodes]{
		\includegraphics[angle=0,scale=.44]{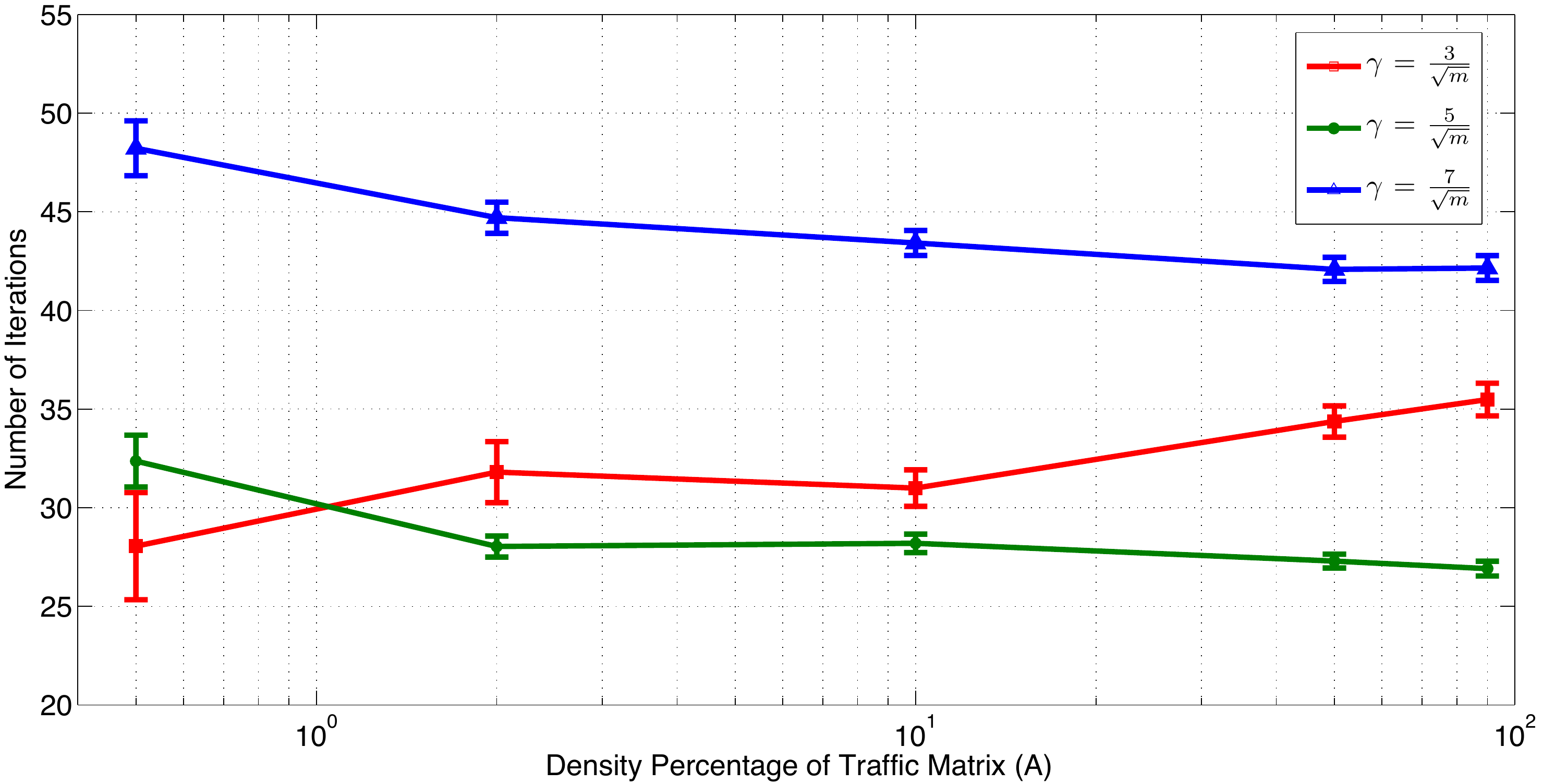}
		\label{fig:N_64_mean}
		}
	\subfigure[Crossbar with $N=128$ nodes]{
		\includegraphics[angle=0,scale=.44]{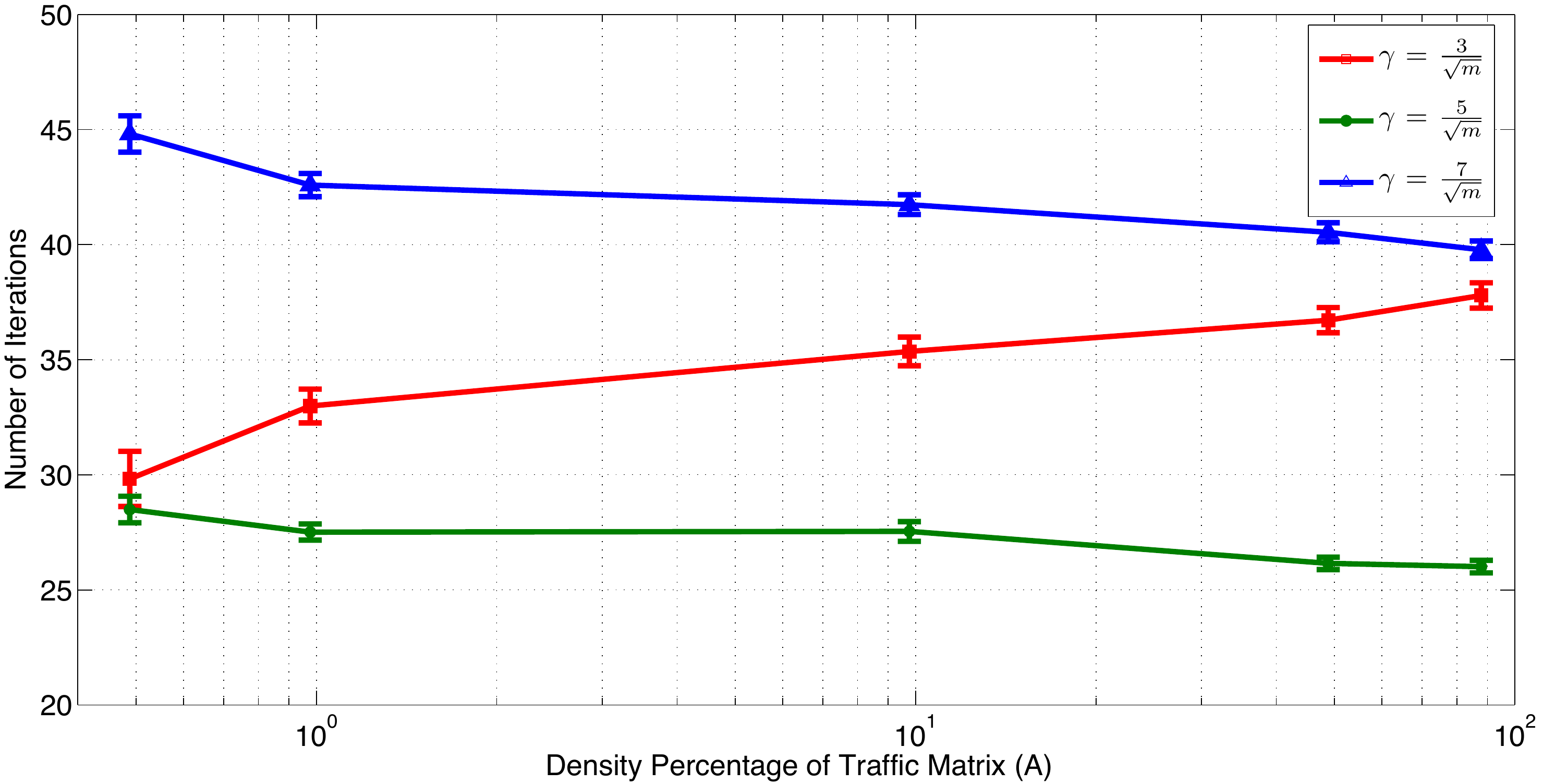}
		\label{fig:N_128_mean}
	}
	\subfigure[Crossbar with $N=256$ nodes]{
		\includegraphics[angle=0,scale=.44]{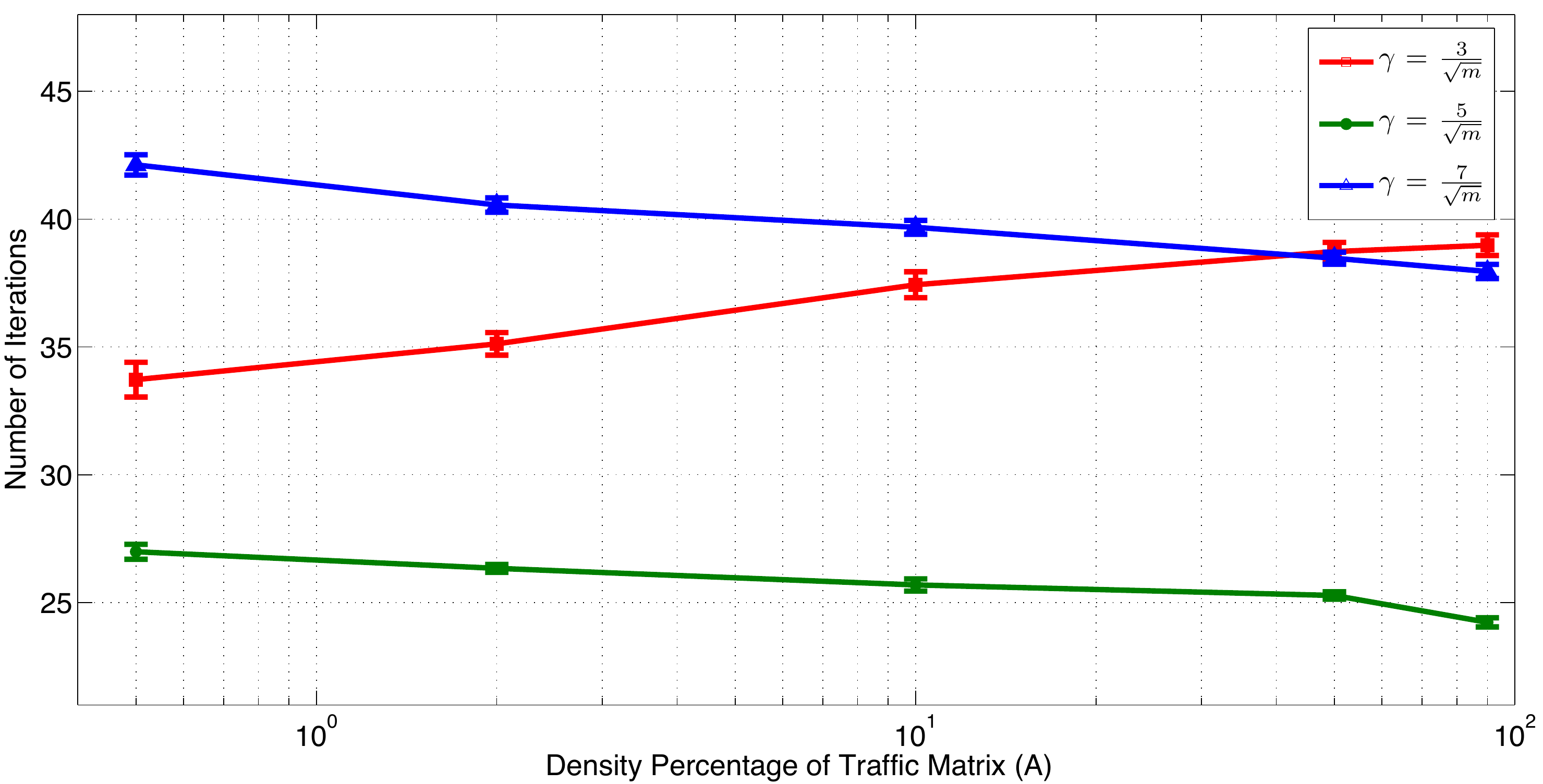}
		\label{fig:N_256_mean}
	}
	\caption{Average number of iterations $\overline I_\epsilon$ for a crossbar with three different number of nodes and $\gamma=\frac{d}{\sqrt{m}}$ for $d=3,5,7$.}
	\label{fig:various_N_mean}
\end{figure}
\begin{table}
	\centering
	\caption{Variance of number of iterations $I_\epsilon$ for different values of $N$, different densities of matrix $\mathbf A$, and step size	$\gamma=\frac{5}{\sqrt{m}}$}
	\begin{tabular}{ | c | c | c | c | c | c |}
		\hline
		\backslashbox[10pt][c]{$N$}{Density of $\mathbf{A}$} & 0.5\% & 2\% & 10\% & 50\% & 90\% \\ \hline
		64 & 44.35 & 7.21 & 5.61 & 3.19 & 3.63 \\ \hline
		128 & 8.49 & 3.17 & 4.93 & 1.97 & 1.91 \\ \hline
		256 & 2.18 & 0.60 & 1.47 & 0.50 & 0.82 \\
		\hline
	\end{tabular}
	\label{tab:table1}
\end{table}

Figure \ref{fig:various_N_mean} portrays  $\overline I_\epsilon$ along with its $90\%$ confidence interval for $N=64, 128,$ and $256$. We see that even with increase in the size of the crossbar $N$, the quantity $\overline I_\epsilon$ increases intangibly. This confirms that Algorithm \ref{alg:iterative_solution} exhibits very good scalability properties, making it useful for many-core systems. Such a tractable scalability stems mainly from salient scalability properties of the diagonally scaled gradient projection algorithm applied in Section \ref{sec:Opt_Sol} to solve the dual problem. The results also imply that depending on $N$ and the density of $\mathbf A$, each of the three chosen step sizes might be preferable. For example, for dense $\mathbf A$, the step size $\gamma=\frac{5}{\sqrt{m}}$ would require fewer iterations. Considering all of the above cases reveals that $\overline I_\epsilon$ diminishes at most to $50\%$ of the choice with the maximum magnitude.
This might sound a dramatic decrease in the average number of required iterations $\overline I_\epsilon$. However, for most cases, such a reduction will be at most about 20-25 iterations.
Table \ref{tab:table1} lists the variance of $I_\epsilon$ for several combinations of traffic matrix densities and crossbar sizes with step size $\gamma=\frac{5}{\sqrt m}$.

Figure \ref{fig:various_N_mean} along with Table \ref{tab:table1} show that confidence intervals gradually shrink as the density of the traffic matrix $\mathbf A$ increases. This is because increasing the density of $\mathbf A$ is equivalent to decreasing the randomness in the pattern of the matrix. Thus, many experiments will encounter (partially) similar traffic patterns for which the $I_\epsilon$ values will be very close. This results in a decrease in the variance of $I_\epsilon$, and hence, causes the confidence interval of $\overline I_\epsilon$ to shrink.

To summarize, the experiments reported above corroborate that (i) Algorithm \ref{alg:iterative_solution} possesses very good and tractable scalability properties, and (ii) it requires only a few tens of iterations (about 25-50) to achieve a solution with high accuracy for a wide range of traffic patterns, choices of step sizes, and number of nodes.
\subsection{Trimming Optimal Rates}
\label{sec:alg2}
In an optical on-chip crossbar, the rate granularity of each data channel is equal to the corresponding rate of one single wavelength $R$. As a result, allocated rates in practical scenarios must be a multiple of $R$. The optimal rates calculated by Algorithm \ref{alg:iterative_solution}, i.e., $\mathbf X^\star$, might not satisfy this property and therefore, some rounding might be necessary. Let $\hat{\mathbf X}$ denote the optimal rate vector after the rounding process. In order to compute $\hat x_{nk}$ from $x^\star_{nk}$, we first decrement $x^\star_{nk}$ to the nearest legal value, and then define $S$ as
$$S=\sum_{n=1}^N\sum_{k=1}^N\left(x^\star_{nk}\; \mathtt{mod}\; R\right).$$
To derive $ \mathbf {\hat X}$, we propose a procedure outlined in Algorithm \ref{alg:trim}. This algorithm consists in choosing $\hat x_{nk}\in\hat{\mathbf X}$ at random with probability proportional to
$x^\star_{nk}-\hat x_{nk}$, and then incrementing $\hat x_{nk}$ by $R$ (provided that feasibility conditions are preserved) and decrementing $S$  by $R$.

\begin{algorithm}[h]
\small
   \caption{Trimming Optimal Rates}
   \label{alg:trim}
\begin{algorithmic}
   \STATE Calculate for all $k$ and $n$:\; $\hat x_{nk}=x^\star_{nk}-\left(x^\star_{nk}\textrm{ }\mathtt{mod}\textrm{ }R\right)$.
   \STATE Calculate $S=\sum_{n=1}^N\sum_{k=1}^N\left(x^\star_{nk}\textrm{ }\mathtt{mod}\textrm{ }R\right)$.
   \WHILE{$S>0$}
    \STATE Let $\mathcal X=\left\{\hat x_{nk}:x^\star_{nk}-\hat x_{nk}>0,\;\sum^N_{n=1}\hat x_{nk}+R\leq \frac{M_k}{\delta}+r_k,\forall n,\forall k\right\}$.
    \STATE Choose an element of $\mathcal X$ randomly with probability proportional to $x^\star_{nk}-\hat x_{nk}$ and set: $\hat x_{nk} = \hat x_{nk} + R$.
	\STATE $S\leftarrow S-R$
   \ENDWHILE
\end{algorithmic}
\normalsize
\end{algorithm}

\subsection{The Case of Bursty Traffic}
We propose another solution procedure which, in contrast to Algorithm \ref{alg:iterative_solution}, is not iterative. Yet it outputs a rate allocation which is optimal in some cases. Such a non-iterative procedure is very fast and  efficient, and thus may prove useful for the case of bursty traffic. This procedure, described in Algorithm \ref{alg:burst_solution}, is motivated as follows. First note that Algorithm \ref{alg:burst_solution} always outputs a feasible point for problem (\ref{eq:opt_prb}). Namely, its output satisfies constraints (\ref{eq:constraint1})-(\ref{eq:constraint2}). Moreover, recall from (\ref{eq:x_opt_alpha}) that for the case of proportionally fair utility functions (i.e., $\alpha=1$), optimal rates can be given by
\begin{align}
\label{eq:x_opt_PF}
x_{nk}^\star=\frac{a_{nk}w_{nk}}{\lambda^\star_k+\lambda^\star_0}.
\end{align}
In order to justify that the output of Algorithm \ref{alg:burst_solution} is a good approximate solution, we consider two cases.

\noindent\textit{Case 1: Under-utilized regime.}
In this case, we assume that under optimal rate allocation, the whole resources of the system are not used by the nodes. Namely, we would have: \mbox{$\sum_{n=1}^N \sum_{k=1}^N x^\star_{nk} < CR$}.
Hence, KKT condition (\ref{eq:KKT_constraint_total}) implies that $\lambda^\star_0=0$, and therefore (\ref{eq:x_opt_PF}) gives
\begin{align}
	\label{eq:x_PF_under}
	x_{nk}^\star=\frac{a_{nk}w_{nk}}{\lambda_k^\star}, \quad \forall k,\forall n.
\end{align}
Hence, we obtain
\begin{align*}
	\frac{a_{n'k}w_{n'k}}{x^\star_{n'k}}=\frac{a_{nk}w_{nk}}{x^\star_{nk}}=\lambda^\star_k, \quad \forall n, n'.
\end{align*}
Now consider receiver node $k$. First note that in this case $\lambda^\star_k>0$ and hence, KKT condition  (\ref{eq:KKT_constraint_k}) implies that
\begin{align*}
	\sum_{n=1}^N a_{nk}x^\star_{nk}=r_k+\frac{M_k}{\delta}.
\end{align*}

Combing these two last relations, we obtain
\begin{align}
	\label{eq:PF_under_optimal}
	x_{nk}^\star=\frac{a_{nk}w_{nk}}{\sum_{j=1}^N a_{jk}w_{jk}}\left(r_k+\frac{M_k}{\delta}\right), \;\; \forall n.
\end{align}
Finally, since the system is under-utilized, we will have $S=CR-\sum_{n=1}^N \sum_{k=1}^N x^\star_{nk}>0$, and therefore the output of Algorithm \ref{alg:burst_solution} is given by (\ref{eq:PF_under_optimal}). Thus, Algorithm \ref{alg:burst_solution} gives the optimal solution for the under-utilized case.

We remark that the solution (\ref{eq:PF_under_optimal}) is intuitive: for any receiver node $k$, the quantity $\sum_{j=1}^N a_{jk}w_{jk}$ is the sum of the weights of the nodes who wish to send packets to node $k$. Thus, (\ref{eq:PF_under_optimal}) implies that the optimal allocation shares the available capacity $r_k+\frac{M_k}{\delta}$ among nodes $(n, \; a_{nk}=1)$  proportionately to their weights.

\noindent\textit{Case 2: Fully-utilized regime.}
In this case, at the optimal point we will have $\sum_{n=1}^N \sum_{k=1}^N x^\star_{nk} = CR$, and so by KKT condition  (\ref{eq:KKT_constraint_total}), we will have $\lambda^\star_0>0$. It then follows from (\ref{eq:x_opt_PF}) that
\begin{align}
	\label{eq:xstar_full}
	x_{nk}^\star=\frac{a_{nk}w_{nk}}{\lambda_k^\star+\lambda_0^\star}< \frac{a_{nk}w_{nk}}{\lambda_k^\star}, \;\; \forall k,\forall n.
\end{align}
Hence, for any $n$ and $k$, if we use the rate given by (\ref{eq:PF_under_optimal}), i.e., choose
\begin{align}
	\label{eq:PF_approx_over}
	x_{nk}=\frac{a_{nk}w_{nk}}{\sum_{j=1}^N a_{jk}w_{jk}}\left(r_k+\frac{M_k}{\delta}\right), \;\; \forall n,
\end{align}
the constraint (\ref{eq:constraint2}) will be satisfied. However, we have that $x_{nk}^\star<x_{nk}$, and since the system is fully-utilized, we will have
\begin{align*}
	CR-\sum_{n=1}^N\sum_{k=1}^N a_{nk}x_{nk}<CR-\sum_{n=1}^N\sum_{k=1}^N a_{nk}x_{nk}^\star=0.
\end{align*}
Thus, constraint (\ref{eq:constraint1}) will be violated (equivalently $S<0$).
To resolve this issue, we consider the uniform rate allocation $\frac{CR}{\sum_{n}\sum_{k} a_{nk}x_{nk}}$ and set $x_{nk}$ to the minimal value between (\ref{eq:PF_approx_over}) and the uniform allocation.
It then follows that this latter choice of $x_{nk}, \forall n,k$ satisfies both constraints. Thus, in this case,	Algorithm \ref{alg:burst_solution} generates a sub-optimal but feasible rate allocation. Finally, we remark that when $\lambda_0^\star \ll \lambda^\star_k, \; \forall k$, this solution is near-optimal as verified by (\ref{eq:xstar_full}).

\begin{algorithm}[h]
\small
   \caption{Solution for Burst Traffic}
   \label{alg:burst_solution}
\begin{algorithmic}
   \STATE 	Compute for all $n$ and $k$: \;
	$
	z_{nk}=\frac{a_{nk}w_{nk}}{\sum_{j=1}^N a_{jk}w_{jk}}\left(r_k+\frac{M_k}{\delta}\right).
	$
   \STATE Compute:\; $S=CR-\sum_{n=1}^N\sum_{k=1}^N a_{nk}z_{nk}.$
   \IF{$S\ge 0$}
    \STATE Set $x_{nk}^\star=z_{nk}$ for all $n,k$.
   \ELSE
    \STATE Set for all $n$ and $k$: \;
	$
	x_{nk}^\star=\min\left(z_{nk},\frac{CR}{\sum_{n=1}^N\sum_{k=1}^N a_{nk}}\right).
	$
   \ENDIF
\end{algorithmic}
\normalsize
\end{algorithm}

\subsection{$\alpha$-Fair Admission Control Algorithm}
\label{sec:alg3}
We now describe the algorithm for optimal wavelength assignment and buffer management based on a central on-chip admission controller. The algorithm is listed below as Algorithm \ref{alg:admission_control_alg}. A built-in admission controller, which from now on we refer to as the controller, is supposed to be mounted onto the system to implement this algorithm. At each time slot $t$, all nodes are required to send their requests to the controller. The request of each node consists of the set of its target destinations, and corresponds to a row in the matrix $\mathbf A$. Moreover, the request contains information about each node's available buffer space as well as the rate at which incoming packets can be processed. After receiving all requests, the controller acquires $a_{nk}$, $w_{nk}$, $M_k$, and $r_k$ for every $n$ and $k$. Then, the controller calculates the optimal rate allocation using Algorithm 1 or 3. According to the final obtained values for rate allocations $\hat{\mathbf X}$, the controller assigns distinct wavelengths on the available waveguides to each node for sending and receiving data.

\begin{algorithm}[h]
\small
   \caption{$\alpha$-Fair Admission Control Algorithm for Buffer Management and Wavelength Assignment}
   \label{alg:admission_control_alg}
\begin{algorithmic}
	\STATE At time slot $t$:
	\STATE \quad Get $M_k^{t+1}$, $r_k^{t+1}$, $a_{nk}^{t+1}$ and $w_{nk}^{t+1}$ for $n,k=1,\dots,N$.
	\STATE \quad Calculate optimal rate values for time slot $t+1$ using Algorithm 1 or 3.
	\STATE Trim the values acquired in Step 2 using Algorithm 2.
	\STATE Based on the rate allocation values from Step 3, inform each node of the channels it should use in time slot $t+1$.
\end{algorithmic}
\normalsize
\end{algorithm}

From computational complexity point of view, the proposed admission controller must be capable of doing simple mathematical and logical operations to implement Algorithm 3. It is worth noting that the output of the controller should be communicated to nodes in a simple yet fast manner.
This may require the system to be equipped with auxiliary optical/electrical connections, as a dedicated signaling media, to communicate admission control results from the controller to nodes as well as delivering requests in the reverse direction. Such a separate media that decouples transmission of data packets and control packets calls for the recently appreciated SDN architectures in networking research community that decouple network control and forwarding functions. This architecture has also been employed in a nanophotonic NoC named 2D-HERT \cite{koohiHERT}.
At each time slot, nodes start sending and receiving data over the crossbar based on the results received from the controller. Meanwhile, nodes send to the controller their requests for the next time slot. Thus, data and control communications can be overlapped so that admission control results are readily accessible by nodes at the beginning of each time slot and controller-related overheads are removed. A similar approach has also been utilized in \cite{Zhao_SDMAC} to increase the channel efficiency of a wireless NoC MAC protocol.

It is worth noting that the overlap between data and control packets can be achieved only if the length of each time slot $\delta$ is greater than or equal to the time it takes to send the requests to the controller and get the results back. This condition determines the minimum required bandwidth and the number of waveguides for communications to/from the controller. In this regard, let \textsf{rqst} and \textsf{rslt} respectively represent the size of the request and result vectors communicated between each node and the controller. Moreover, let $B_c$ denote the bandwidth of the channel between each node and the controller. Then, the required time for sending the requests to the controller and getting the results back would be at most $\frac{\textsf{rqst}}{B_c}+\frac{\textsf{rslt}}{B_c}+V$, where $V$ is the maximal overhead due to controller's computations. Thus, we require
\begin{align}
\label{eq:B_c_LB1}
\frac{\textsf{rqst}}{B_c}+\frac{\textsf{rslt}}{B_c}+V\leq \delta, \end{align}
or equivalently: $B_c\geq \frac{\textsf{rqst}+\textsf{rslt}}{\delta-V}.$

From (\ref{eq:B_c_LB1}), we can also compute the total number of waveguides required to realize the communication channels between the central controller and the on-chip nodes. For instance, in a crossbar with $N$ nodes and 64 wavelengths with rate $R$ multiplexed on each waveguide, the number of waveguides $W_c$ should satisfy
\begin{align}
	\label{eq:Wc}
	W_c\geq \lceil\frac{NB_c}{64 R}\rceil\geq\lceil\frac{N(\textsf{rqst}+\textsf{rslt})}{64 R(\delta-V)}\rceil.
\end{align}
\section{Hardware Implementation}
\label{sec:HwImp}
In this section a practical implementation of the proposed controller is described. It is followed by an analysis of its area and power consumption overheads.
\subsection{Hardware Design}
\begin{figure*}
	\centering
	\subfigure[Low latency implementation]{
		\includegraphics[angle=0, scale=0.65]{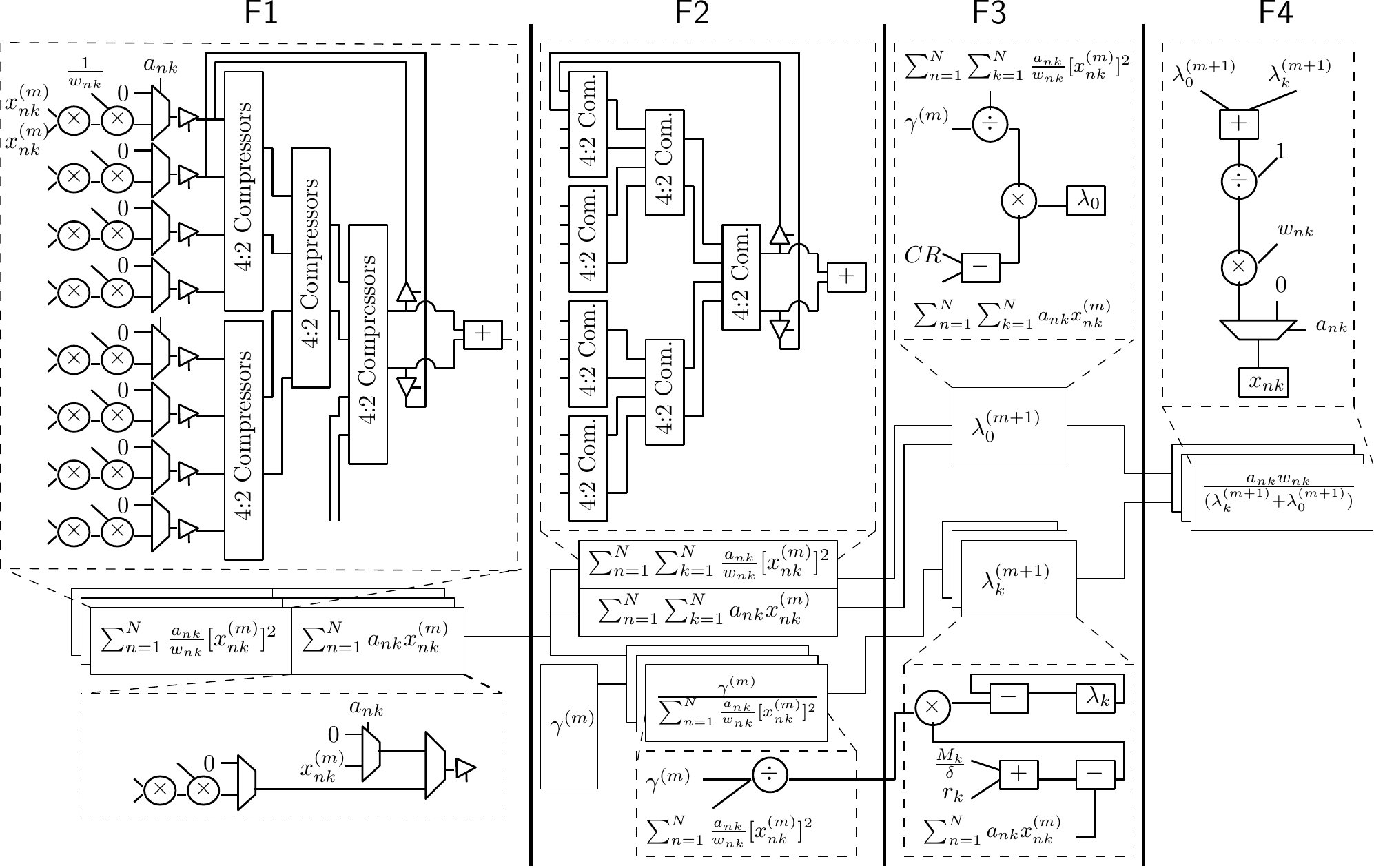}
		\label{fig:basicStructureHighPerf}
	}
	\vspace{4mm}
	\hrule
	\vspace{4mm}
	\subfigure[Low power implementation]{
		\includegraphics[angle=0,scale=0.68]{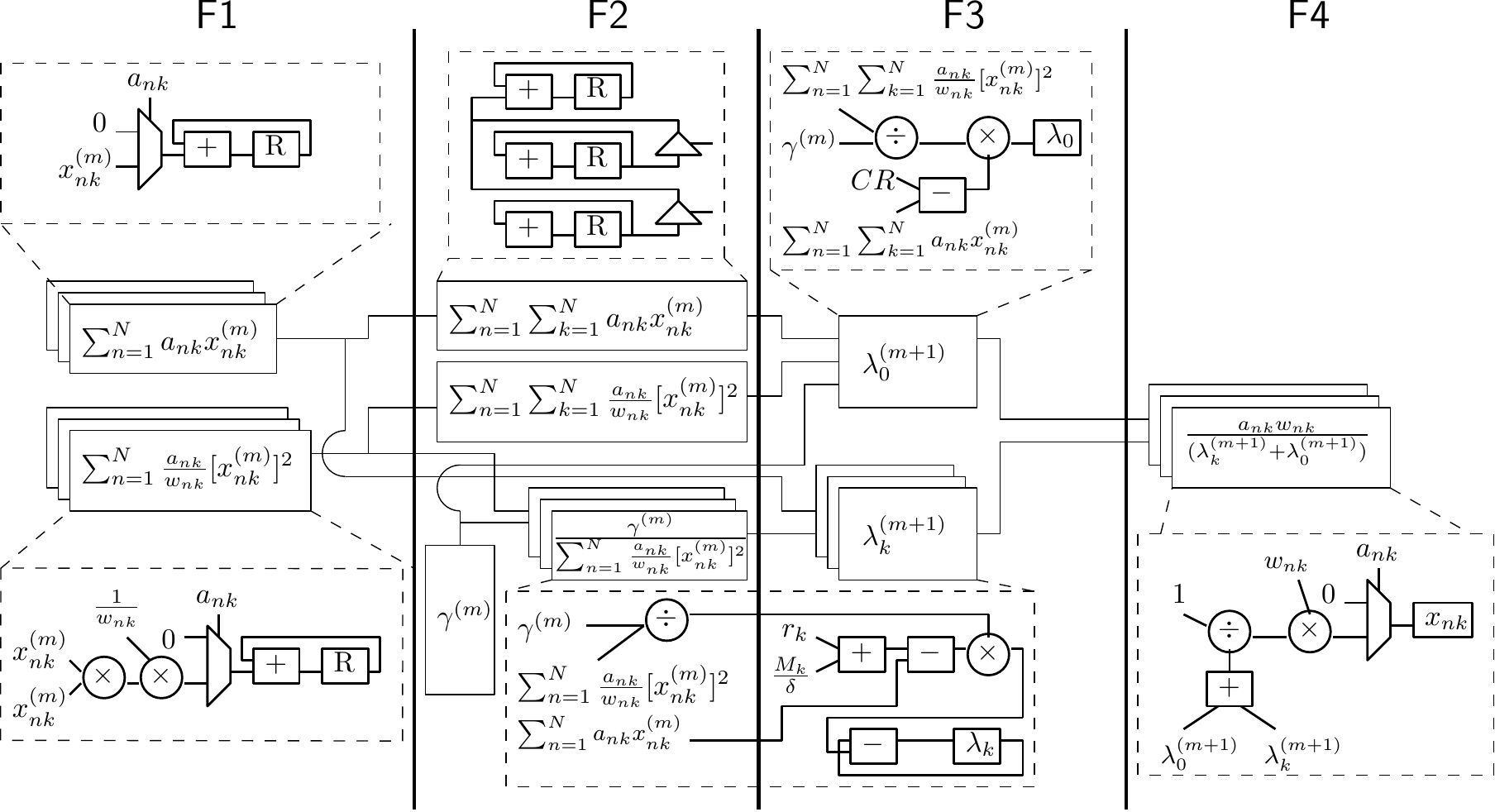}
		\label{fig:basicStructureLowPow}
	}
	\caption{The basic structure of different implementations of the proposed controller.}
	\label{fig:basicStructre}
\end{figure*}
In order to account for both power and latency limitations as the key concerns for the design of nowadays digital systems, we propose two variants: A \emph{low latency} implementation that strives to output the results with minimal latency, and a \emph{low power} implementation  whose aim is consume the least possible power. Figure \ref{fig:basicStructre} portrays the basic structure of the proposed implementation for both variants.	To facilitate the presentation, the concentration in the subsequent design will be on implementation of an iterate of Algorithm \ref{alg:iterative_solution}, thereby ignoring Algorithm \ref{alg:trim}\hbox{  }\footnote{The overhead due to Algorithm \ref{alg:trim} is ignored as an alternative of this algorithm can be simply implemented using shift operation and XOR-based pseudo-random number generation. Furthermore, for practical purposes we propose to run Algorithm \ref{alg:iterative_solution} for a fixed number of iterations, in contrast to the conditional stopping.}. To alleviate implementation complexity, we use fixed point representation for real numbers. In Algorithm \ref{alg:iterative_solution}, each iteration $m$ consists of three basic steps:
\begin{enumerate}
	\item[(i)] Setting step size $\gamma^{(m)}$
	\item[(ii)] Updating dual variables $\lambda_0^{(m+1)}$, $\lambda_k^{(m+1)}$, for all $k = 1,\dots,N$
	\item[(iii)] Calculating primal variables $x_{nk}^{(m+1)}$ for all $n,k$ such that $a_{nk}=1$
\end{enumerate}

Next we describe how to implement these three steps in detail.

\paragraph{Step (i)}
This step, which has the same implementation for both variants, is implemented using a look-up table. Namely, we calculate all values of step size at design time and store them in look-up table array.

\paragraph{Step (ii)}
Computation of this step is divided into three phases denoted by \textsf{F1}, \textsf{F2}, and \textsf{F3} in Figure \ref{fig:basicStructre}. In phase \textsf{F1}, to achieve a high performance controller for low latency implementation, the terms $\sum_{n=1}^N \frac{a_{nk}}{w_{nk}} [x_{nk}^{(m)}]^2$ and $\sum_{n=1}^N a_{nk} x_{nk}^{(m)}$ for $k=1,\dots,N$ are computed in a fully parallel structure. To this end, we first calculate $\frac{a_{nk}}{w_{nk}}[x_{nk}^{(m)}]^2$ and $a_{nk}x_{nk}$ for all $n,k$, and then accumulate the calculated numbers of each column.

Recall that $a_{nk}\in\{0,1\}$ and we simply calculate $a_{nk}x_{nk}^{(m)}$ by a multiplexer. Moreover, to reduce computation complexity of this phase, we store both of $\frac{1}{w_{nk}}$ and $w_{nk}$ ~\footnote{Indeed we require each node $n$ to report both $w_{nk}$ and $\frac{1}{w_{nk}}$, for $k=1,\dots,N$.}. Thus, we can compute $\frac{1}{w{nk}}[x_{nk}^{(m)}]^2$ by using one multiplier and two multiplexers, and control the multiplier inputs by the controller part of this circuit. Finally, for each column $k$, we accumulate different numbers $\frac{a_{nk}}{w_{nk}} [x_{nk}^{(m)}]^2$ for $n=1,\dots,N$ in parallel. To attain high performance circuit for this accumulation, we use a combination of carry save idea \cite{Parhami1999ComArith} and tree structure summation as shown in Figure \ref{fig:basicStructre}. Moreover, sharing this circuit for computations of $\sum_{n=1}^N \frac{a_{nk}}{w_{nk}} [x_{nk}^{(m)}]^2$ and $\sum_{n=1}^N a_{nk} x_{nk}^{(m)}$, the hardware overhead of the proposed controller will be significantly reduced. To this end, the proposed implementation computes summation $\sum_{n=1}^N a_{nk} x_{nk}^{(m)}$ when it calculates $\frac{a_{nk}}{w_{nk}}[x_{nk}^{(m)}]^2$.

Unlike the low latency variant, in phase \textsf{F1} of the low power variant, the terms $\sum_{n=1}^N a_{nk} x_{nk}^{(m)}$ and $\sum_{n=1}^N \frac{a_{nk}}{w_{nk}} [x_{nk}^{(m)}]^2$ for $k = 1,\dots,N$ are computed in partially parallel manner. As a tradeoff between achieving higher performance and lower area and power overhead, we use a pipeline structure similar to multiple and accumulate operation in phase \textsf{F1}. In other words, implementation of this variant performs computation of $\frac{1}{w{nk}}[x_{nk}^{(m)}]^2$ and the summation operation simultaneously. Furthermore, to obtain acceptable performance in low power variant we compute $\sum_{n=1}^N \frac{a_{nk}}{w_{nk}} [x_{nk}^{(m)}]^2$ and $\sum_{n=1}^N a_{nk} x_{nk}^{(m)}$ in parallel using distinct circuits.

Two types of computations are accomplished in phase \textsf{F2}: summation ($\sum_{n=1}^N\sum_{k=1}^N a_{nk} x_{nk}^{(m)}$, $\sum_{n=1}^N\sum_{k=1}^N \frac{a_{nk}}{w_{nk}} [x_{nk}^{(m)}]^2$) and division ($\frac{\gamma^{(m)}}{\sum_{n=1}^N \frac{a_{nk}}{w_{nk}} [x_{nk}^{(m)}]^2}$). For summation, results of phase \textsf{F1} are accumulated. To mitigate hardware overhead of low latency implementation, we use the summation circuit used in phase \textsf{F1} to accumulate $N$ different numbers. However, in low power implementation to attain high performance controller, we use tree structure summation to accumulate $N$ different numbers. Moreover, to mitigate hardware overhead of low power variant, we share the required adders of this computation with adders of phase \textsf{F1} of this variant. For division, computation of all columns is performed simultaneously for both variants. As in \cite{Ercegovac2000Div}, we use multiplier based divider, which is popular in commercial processors. Hence, sharing required multipliers of division with multipliers of phase \textsf{F1} further reduces hardware overhead. Finally, the circuit of phase \textsf{F3} performs last computations of updating dual variables in parallel for both implementations. Again, to alleviate the hardware cost, we share the computational resources of this phase with the computational resources of phase \textsf{F1}.

\paragraph{Step (iii)}
Last phase of each iterate in Algorithm \ref{alg:iterative_solution} is devoted to the calculation of primal variables using dual variables. To this end, we first calculate the denominator of primal variables using dual variables and then compute all primal variables in a fully (resp.~partially) parallel way for low latency (resp.~low power) variant. The basic structure of implementation of this phase is presented as phase \textsf{F4} in Figure \ref{fig:basicStructre}. To obtain a implementation with lower hardware overhead in low latency variant, we reuse multipliers of phase \textsf{F1} in the division and multiplication computation of this phase. It is worth noting that by performing proper resource sharing among different phases in both implementations, the controller part of the implementation circuit controls the inputs of shared resources in datapath using multiplexers and selects correct inputs based on its state. As a tradeoff between achieving higher performance and lower hardware overhead for low power implementation, we calculate the elements of all columns of primal variables in parallel so that elements of each columns are computed using a pipeline approach.
\subsection{Area Overhead}
To have an intuition for the area overhead of the proposed controller, we consider a system with 64 cores, i.e., $N = 64$ in 32 $\mathrm{nm}$ process. To calculate storage overhead, we use CACTI \cite{CACTI}. Moreover, we estimate the computational overhead using the data provided \cite{li2013mcpat} and \cite{Kashfi2008mac} for computational operations. In this system, the total area overhead of the proposed implementation are 9.416539 $\mathrm{mm}^2$ and 1.087179$\mathrm{mm}^2$ for low latency and low power implementations, respectively. In particular, our calculations show that only around 9.326724/662 = 1.5\% of the total chip area (e.g., in the case of Xeon E5-2699 v3) for low latency implementation will be occupied by the controller. Similarly, the area overhead for low power implementation is nearly 1.087179/662 = 0.2\% of the total chip area, signifying that the area overhead of the proposed controller in the case of both variants is negligible.
\subsection{Power Consumption Overhead}
We now compute the power overhead of the controller. Unless stated otherwise, we consider a 32 $\mathrm{nm}$ process.

\paragraph{Dynamic power:} Similarly to \cite{mukundan2012morse}, to estimate the dynamic power consumption of the proposed controller, we use the method presented in McPAT tool \cite{li2013mcpat}. To this end, we count the number of operations involved in each component of the presented implementation. Then, by multiplying this number by the energy consumption of each component that is consumed for each access, we obtain its dynamic energy consumption. Note that we adopt the dynamic energy consumption of computational component (such as multiplier, adder, etc.) and storage component using \cite{Kashfi2008mac} and CACTI \cite{CACTI}, respectively. Let $N = 64$ and assume that each iteration lasts for 27 (resp.~150) clocks for low latency (resp.~low power) implementation. Hence, the total dynamic power consumption of the proposed controller is 6.99 $\mathrm{W}$ and 1.45 $\mathrm{W}$ for low latency and low power variants, respectively.

\paragraph{Leakage Power:}
Using CACTI, we estimate the leakage power of storage parts of the proposed controller. Moreover, using McPAT together with data provided in \cite{Kashfi2008mac}, we calculate the leakage power of computational parts of the controller. We obtain that the total leakage power of the controller for low latency and low power variants are 4.96 $\mathrm{W}$ and 0.25 $\mathrm{W}$, respectively. Finally, we find out that the power overhead of the proposed controller constitutes about 11.95/145 = 8.2\% and 1.7/145 = 1.2\% of the total power consumed by the chip (e.g., Xeon E5-2699 v3) for low latency and low power implementations, respectively. This highlights that the power overhead of the controller especially for low power implementation is negligible.
%
%
\section{Simulation Experiments}
\label{sec:simulation}
For simulation experiments, we use OMNeT++ \cite{Omnet} to simulate an optical MWMR on-chip crossbar equipped with our proposed admission control policy. We will refer to such an architecture as MWMR-AC. We consider an MWMR-AC consisting of 64 nodes, each operating at 5 GHz clock speed. We also consider 64 waveguides for the data channels of the crossbar. Moreover, DWDM technique is used to carry 64 wavelengths on each waveguide simultaneously, providing a total of 4096 data channels. Each wavelength is considered to provide a bandwidth of $R=10$ Gbps. Finally, in all the experiments, we focus on the case of $\alpha=1$, which corresponds to \emph{proportional fairness} metric \cite{Mo2000utility}.

In all experiments we assume $\delta= 6$ ns. This choice of $\delta$ follows from our estimate of controller's minimal delay (5.4 ns) obtained from our hardware implementation in Section \ref{sec:HwImp}. Moreover, based on (\ref{eq:Wc}) and in accordance to our simulated crossbar, we consider 20 waveguides to realize the communication channels between the central controller and the on-chip nodes.

For the sake of comparison, we conduct the experiments with the Corona crossbar \cite{Vantrease-Corona}, with 64 crossbar waveguides each carrying 64 wavelengths simultaneously. Thus, each home node will have one dedicated waveguide (64 wavelengths) for data reception. We also consider Corona with the enhanced ``Fast Forward" token arbitration mechanism (Corona-FF) proposed in \cite{vantrease2009light}.

We consider both synthetic and real traffic patterns in our simulation experiments. In particular, we use Uniform and Hot Spot for synthetic patterns, whereas SPLASH-2 and PARSEC are used for real benchmark evaluations.
\subsection{Synthetic Patterns}
In this subsection, we evaluate the performance of MWMR-AC for two synthetic patterns. In particular, we use the following three measures:
\begin{enumerate}
	\item[(i)] \emph{Latency} defined as the difference between the time a packet arrives at the destination and the time it was generated. We focus on the average latency experienced by all delivered packets.
	\item[(ii)] \emph{Network Throughput} that is the rate at which packets are delivered by the underlying network. We report the aggregate throughput normalized by the total crossbar capacity.
	\item[(iii)] \emph{Nodal Throughput} defined as the rate at which each node can send packets to other destination nodes. Network throughput is indeed the sum of all nodal throughputs across the crossbar.
\end{enumerate}
In all cases, latency and throughput are measured as a function of the offered load injected to the network. Note that offered load values are normalized with respect to the total capacity provided by the crossbar.  An offered load value of 1 represents the maximum load that can potentially be delivered by the crossbar.

\begin{figure}
	\centering
	\subfigure[Average latency; uniform traffic]{
		\includegraphics[angle=0,scale=0.4]{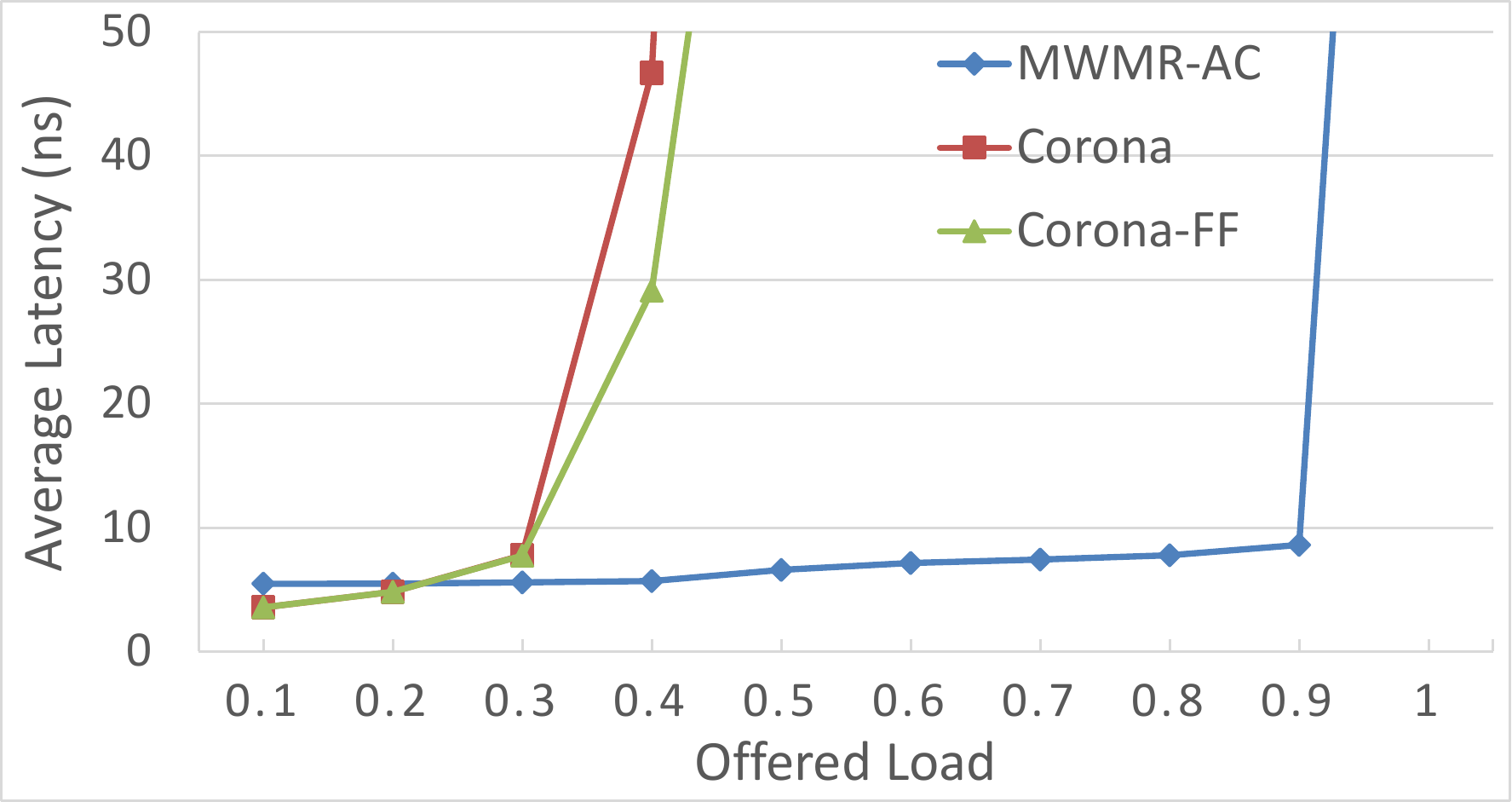}
		\label{fig:uniformL}
	}
	\subfigure[Network throughput; uniform traffic]{
		\includegraphics[angle=0,scale=0.4]{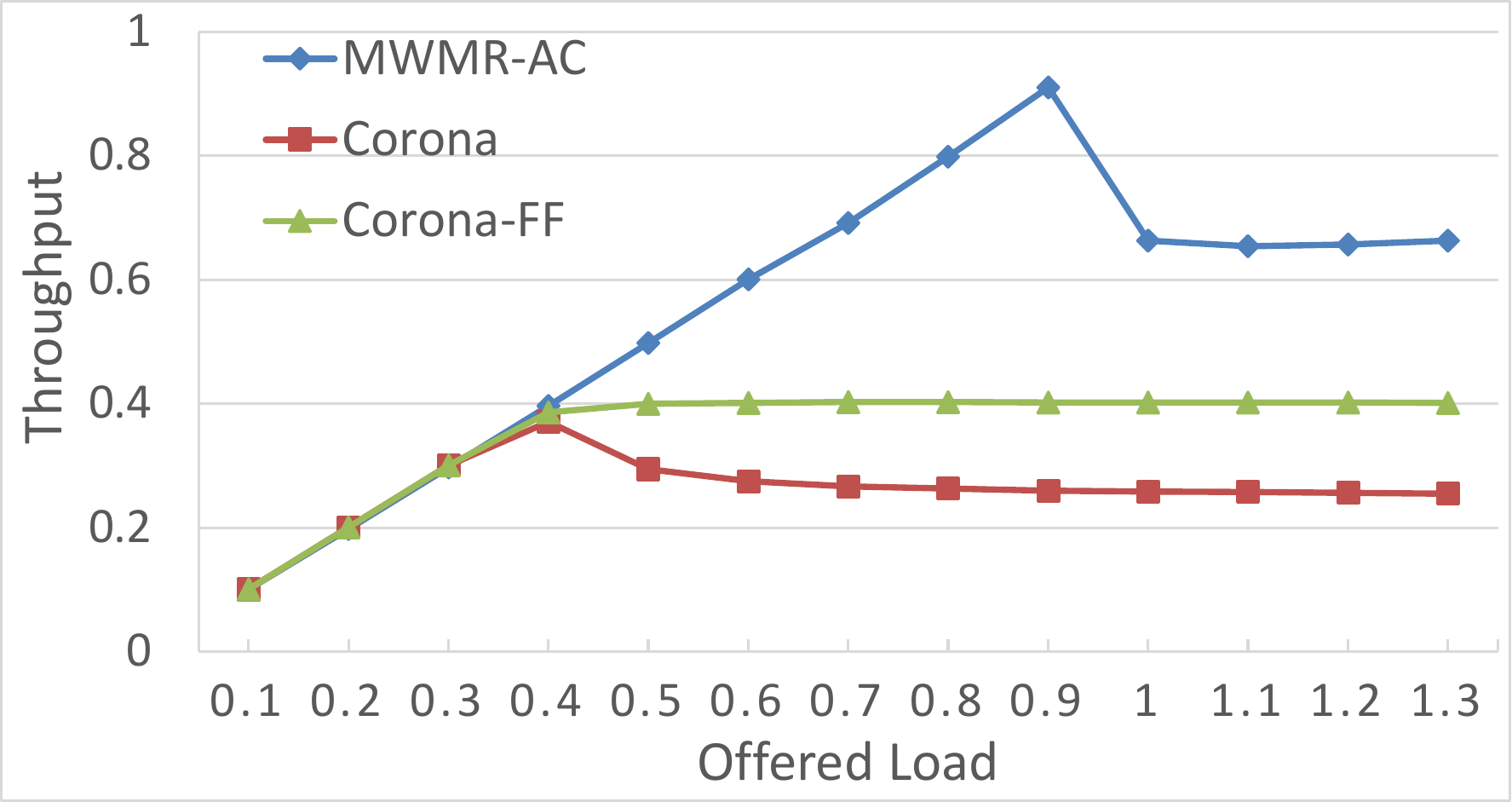}
		\label{fig:uniformT}
	}
	\caption{Average latency and network throughput for the Uniform traffic pattern.}
	\label{fig:uniform}
\end{figure}

Figure  \ref{fig:uniformL} and Figure \ref{fig:uniformT} respectively show the latency and throughput results under the Uniform traffic pattern. Under the Uniform pattern, packet destinations are chosen randomly with a uniform distribution. As shown in Figure \ref{fig:uniformL}, MWMR-AC achieves lower latency than Corona/Corona-FF for offered loads above 0.2. We also see a mild increase in latency up to offered load 0.9 with MWMR-AC, whereas Corona's latency becomes unbounded above offered load 0.4. This is in line with the results shown for throughput in Figure \ref{fig:uniformT}. While MWMR-AC provides a maximum of 0.9 throughput, Corona/Corona-FF can only achieve a maximum throughput of 0.4. Moreover, MWMR-AC hits the saturation point at offered load 1.0, whereas Corona/Corona-FF is saturated at offered load 0.5. The saturation throughput itself ($\approx 0.7$) is also higher for MWMR-AC (0.7) compared to Corona/Corona-FF. We also see a higher saturated throughput for Corona-FF (0.4) compared to Corona (0.3). The Fast Forward token mechanism used in Corona-FF helps Corona to achieve a relatively higher throughput.

Another important observation from Figure \ref{fig:uniformT} is the throughput decrease after offered load 0.9. This is the point where the controller starts using the iterative solution for channel assignments, i.e., it switches from Algorithm \ref{alg:burst_solution} to Algorithm \ref{alg:iterative_solution}. Consequently, the higher delay of the iterative solution causes about 0.2 drop in throughput. However, throughput is not decreased any further and is saturated around 0.7.

Latency results also show that for offered loads 0.1 and 0.2, Corona/Corona-FF provides slightly lower latency. This is due to the fact that with such low traffic loads, controller's delay becomes the dominant latency factor in MWMR-AC. In such cases, demand for communication resources is too low that prevents us from achieving any benefits from better sharing and efficient utilization of channels.
\begin{figure}
	\centering
	\subfigure[Average latency; Hot Spot traffic]{
		\includegraphics[angle=0,scale=0.52]{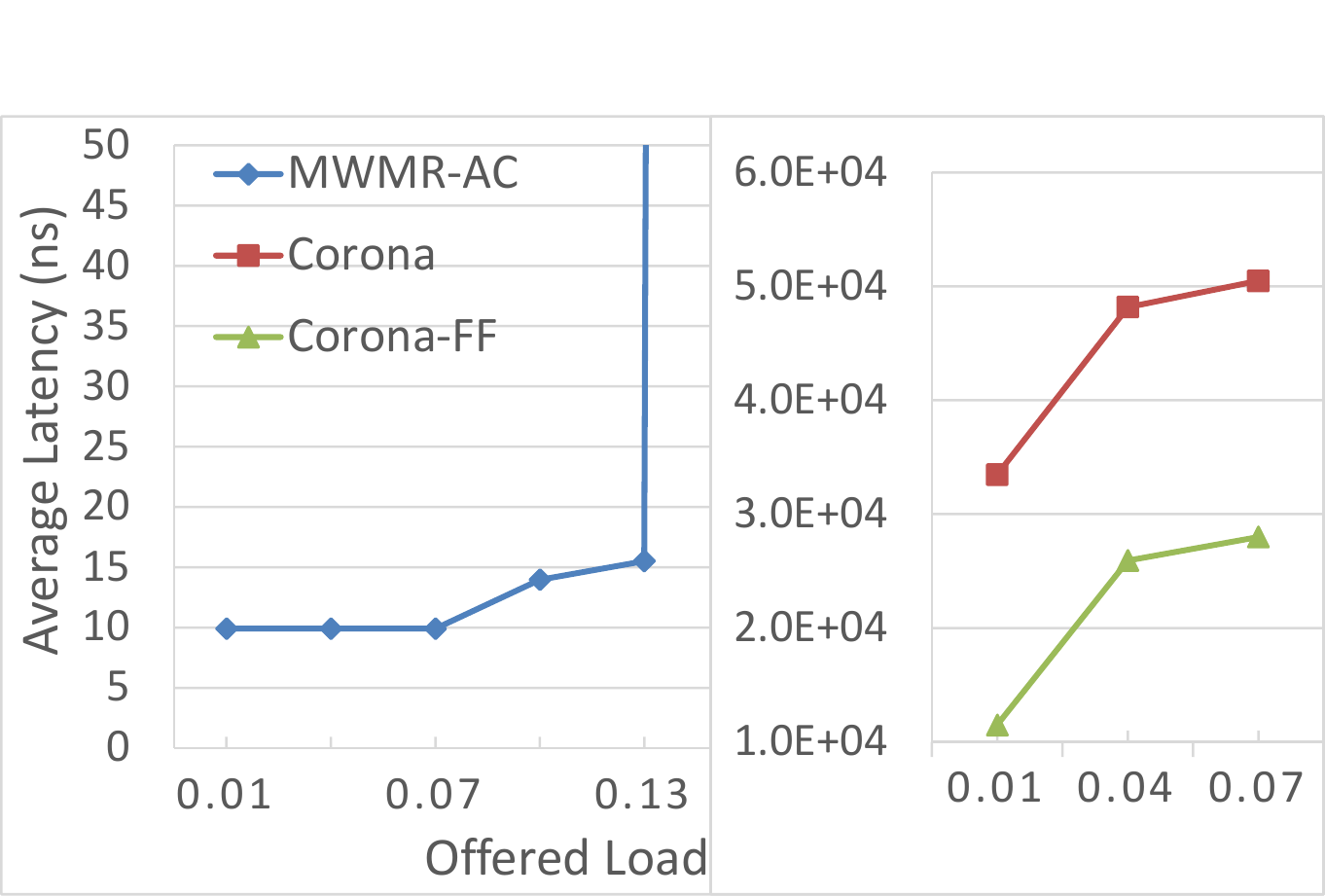}
		\label{fig:hotSpotL}
	}
	\subfigure[Network throughput; Hot Spot traffic]{
		\includegraphics[angle=0,scale=0.52]{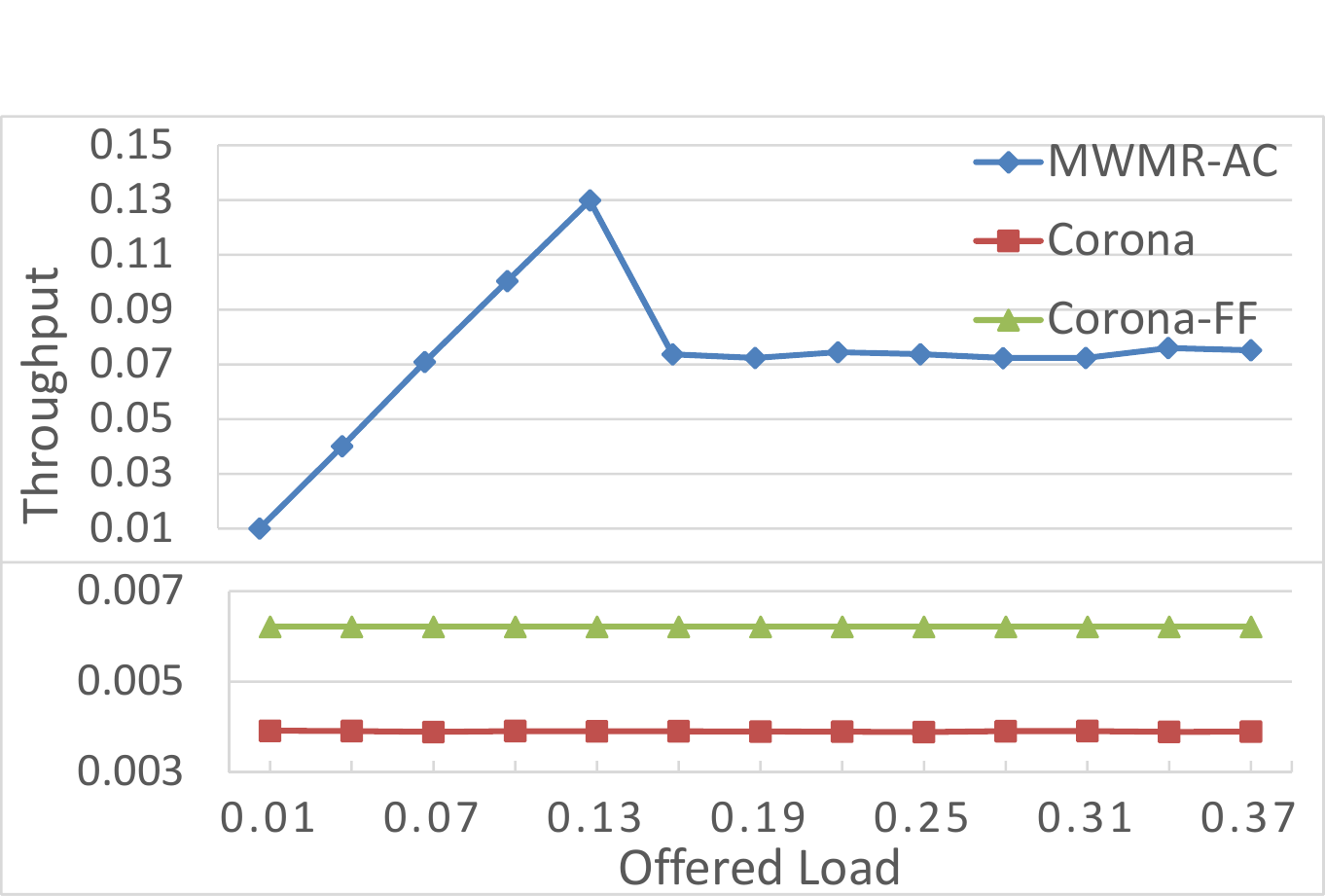}
		\label{fig:hotSpotT}
	}
	\caption{Average latency and network throughput for the Hot Spot traffic pattern.}
	\label{fig:hotspot}
\end{figure}

Figure \ref{fig:hotspot} shows the latency and throughput results for the Hot Spot pattern. Under the Hot Spot pattern, all nodes send their packets to one single destination node. Without loss of generality, here we assume that node 0 is the Hot Spot destination. Latency results in Figure \ref{fig:hotSpotL} show that MWMR-AC achieves a significantly lower latency compared to Corona/Corona-FF in the Hot Spot pattern. Note the different scale used in the two latency axes. In the case of MWMR-AC, the average latency is 3 orders of magnitude lower. This is because MWMR-AC has a much better sharing and utilization of all the channels that are provided by the crossbar. With the Hot Spot pattern, all nodes attempt to access the limited number of channels that are provided by the single waveguide that is dedicated to node 0 for data reception. The resulting high contention will also increase the overheads of the token-based arbitration mechanism used in Corona as token of node 0 will be without any credit most of the time. On the contrary, the admission control policy in MWMR-AC utilizes all the data channels available in the crossbar, and efficiently assigns them to the set of source nodes.

From Figure \ref{fig:hotSpotT}, we can also see that MWMR-AC achieves higher throughput compared to Corona/Corona-FF. While Corona/Corona-FF is already at the saturated point from the initial 0.01 offered load, we see linear increase in the throughput of MWMR-AC up to 0.13. After that, similarly to what we discussed for the case of Uniform pattern, the throughput is decreased to a saturated bound of 0.07. As shown, this is about 10 times higher than the saturation throughput of Corona/Corona-FF which is less than 0.007. It is worth mentioning that the maximum throughput achieved under the Hot Spot traffic is lower than that of Uniform traffic. This is totally expected; the limited buffer space and drain rate of the single destination node (i.e., node 0) limits the total achievable bandwidth for the Hot Spot pattern.
\subsection{Fairness Analysis}
In another experiment, we evaluate the behavior of our proposed admission control policy in terms of fairness. To this aim, we measure the nodal throughput experienced by each node for sending packets. This result highlights the portion of the capacity of crossbar allocated to each node. The Hot Spot traffic pattern with a high offered load value is used for these experiments since fairness issues mainly arise in the presence of contention for resources. We note, however, that our results with high-load Uniform traffic pattern (not shown here) exhibit a similar trend.

We first enumerate nodes from 0 to 63, and distribute them into three categories. In particular, nodes 0 to 20 belong to the first category, nodes with indices 21 to 41 belong to the second category, and the rest belong to the third category. Each category is then assigned with a weight so that all nodes in the same category will have the same weight for sending data to other receiving nodes.

We consider two scenarios in terms of the weights assigned to the three categories. In the first scenario, all categories will have the same weight (equal to 1), whereas in the second one, the weight for the first, the second, and the third category of nodes is respectively set to 1, 2, and 3.

\begin{figure}[ht]
	\centering
	\subfigure[MWMR-AC; different weights]{
		\includegraphics[angle=0,scale=.405]{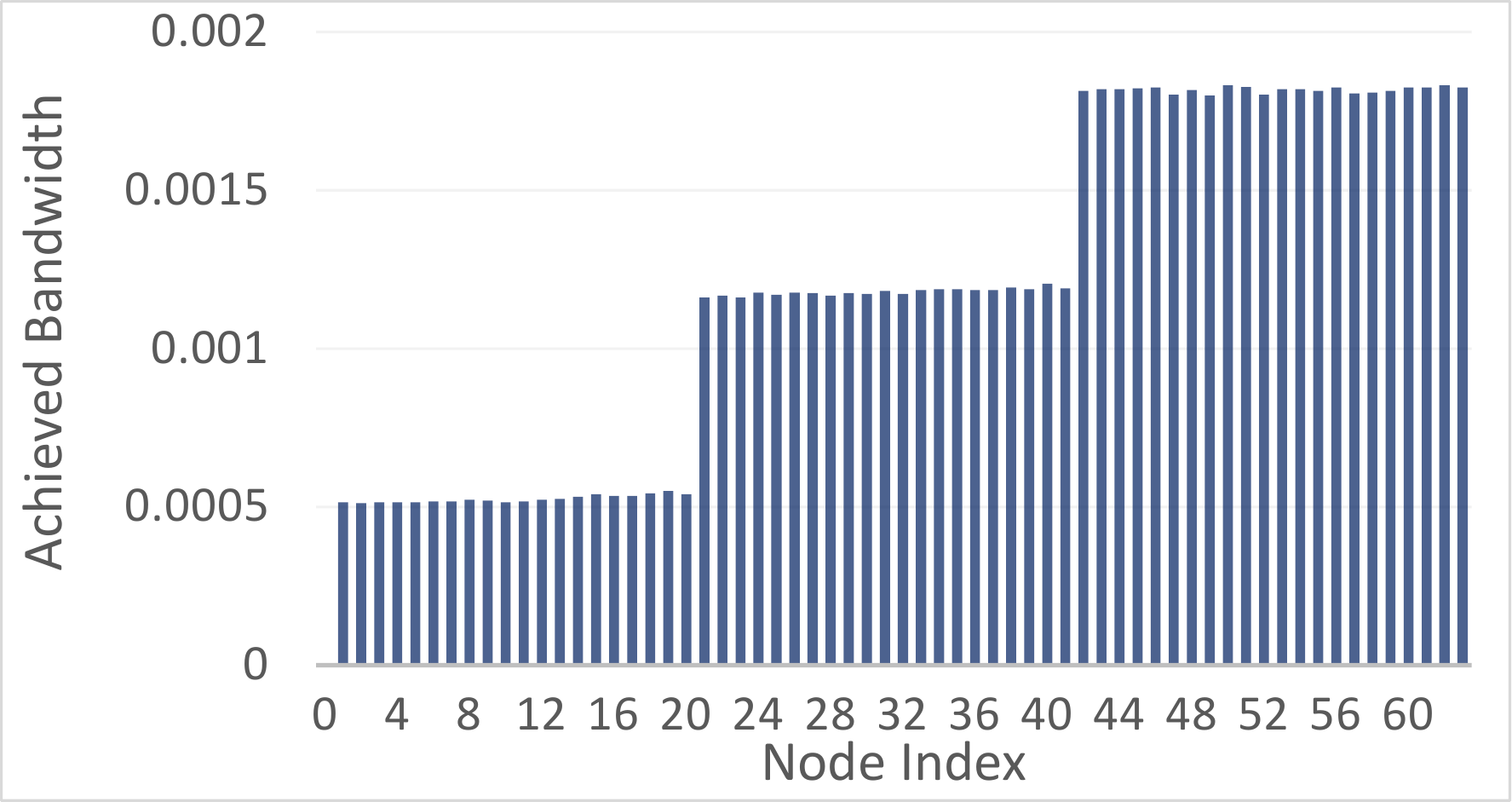}
		\label{fig:ACHotspotIndDiffW}
	}
	\subfigure[MWMR-AC; equal weights]{
		\includegraphics[angle=0,scale=.405]{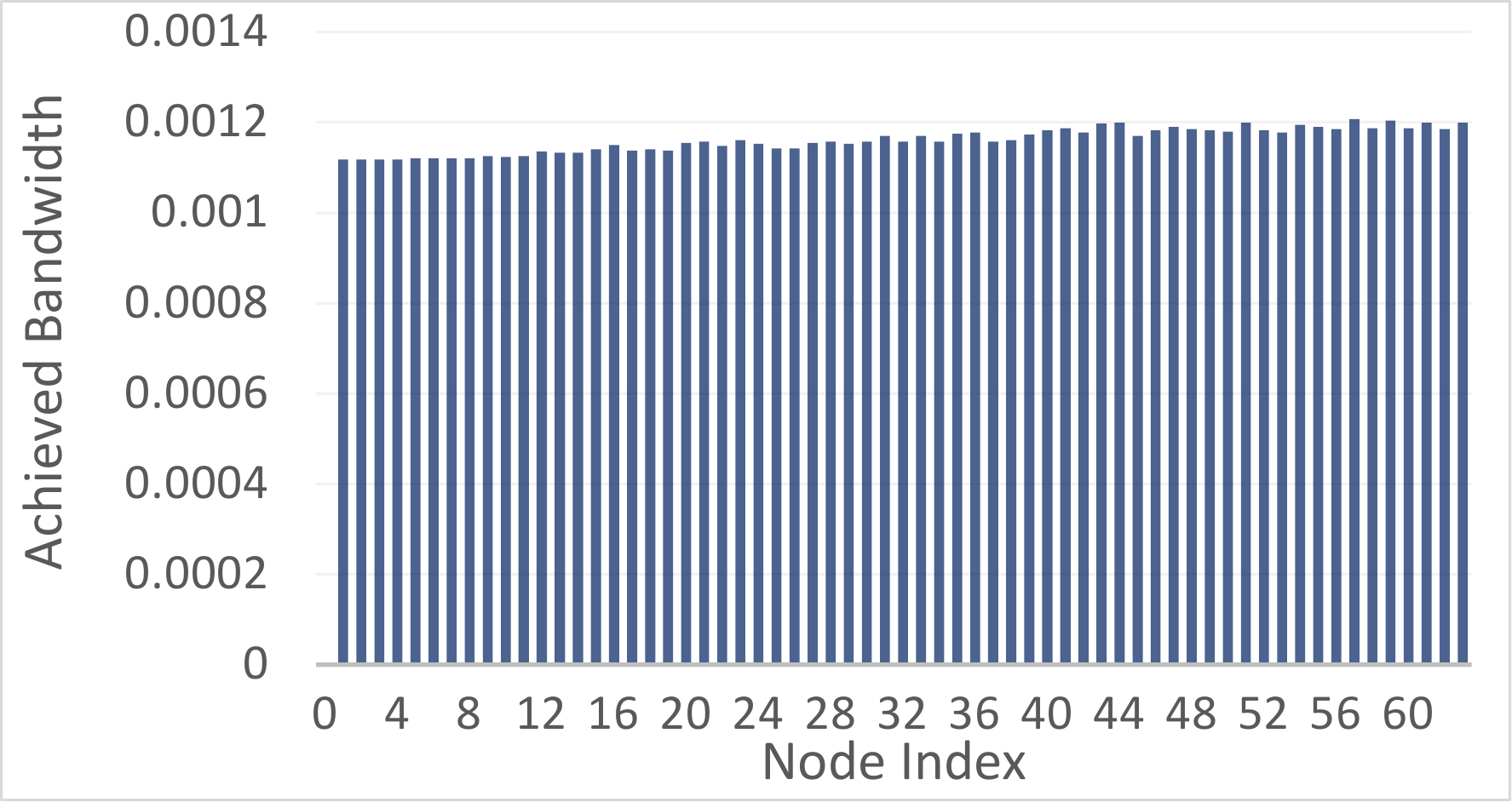}
		\label{fig:ACHotspotIndSameW}
	}\\
	\subfigure[Corona; equal weights]{
		\includegraphics[angle=0,scale=.405]{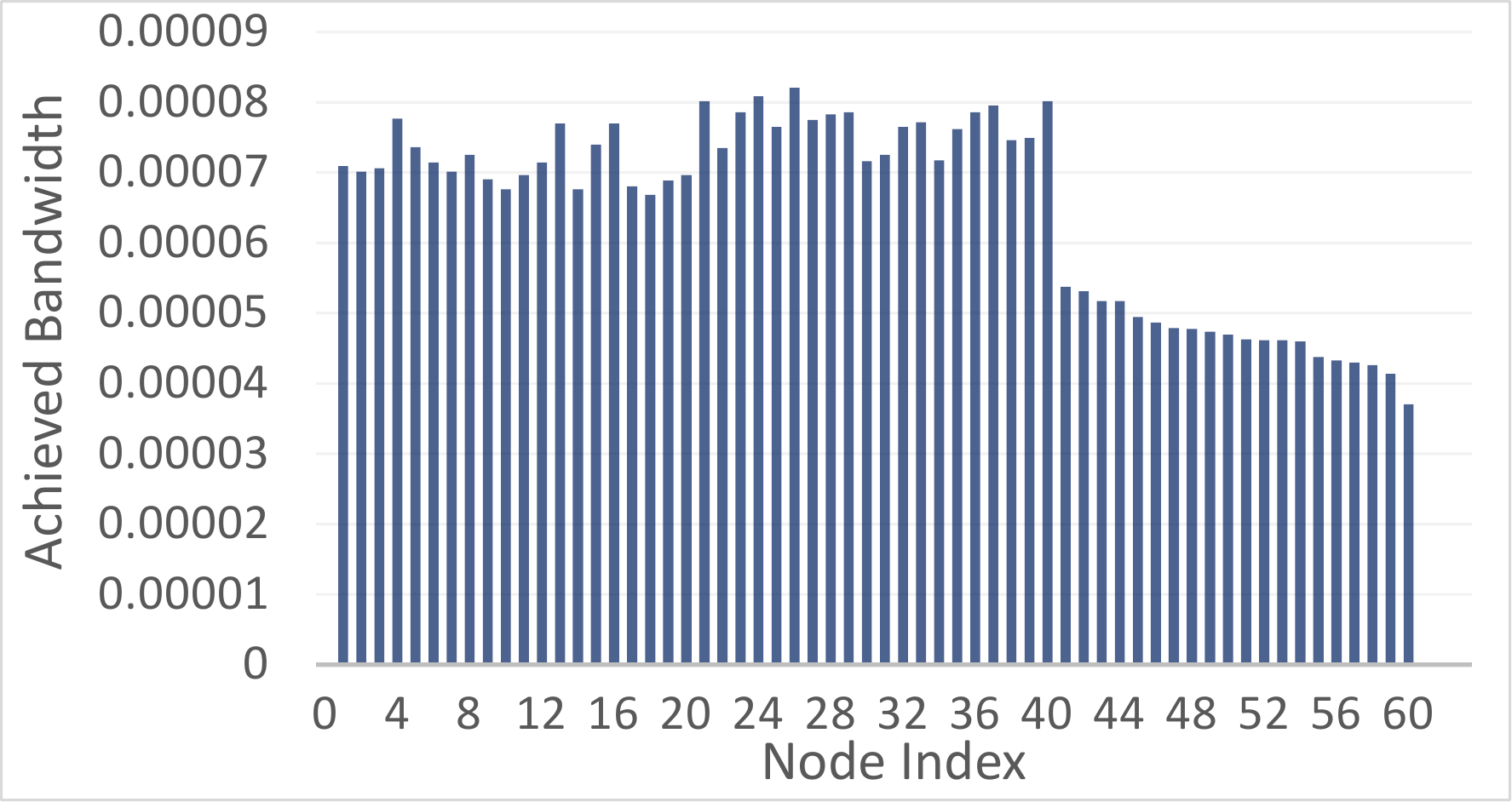}
		\label{fig:noFFHotspotInd}
	}
	\subfigure[Corona with FF token; equal weights]{
		\includegraphics[angle=0,scale=.405]{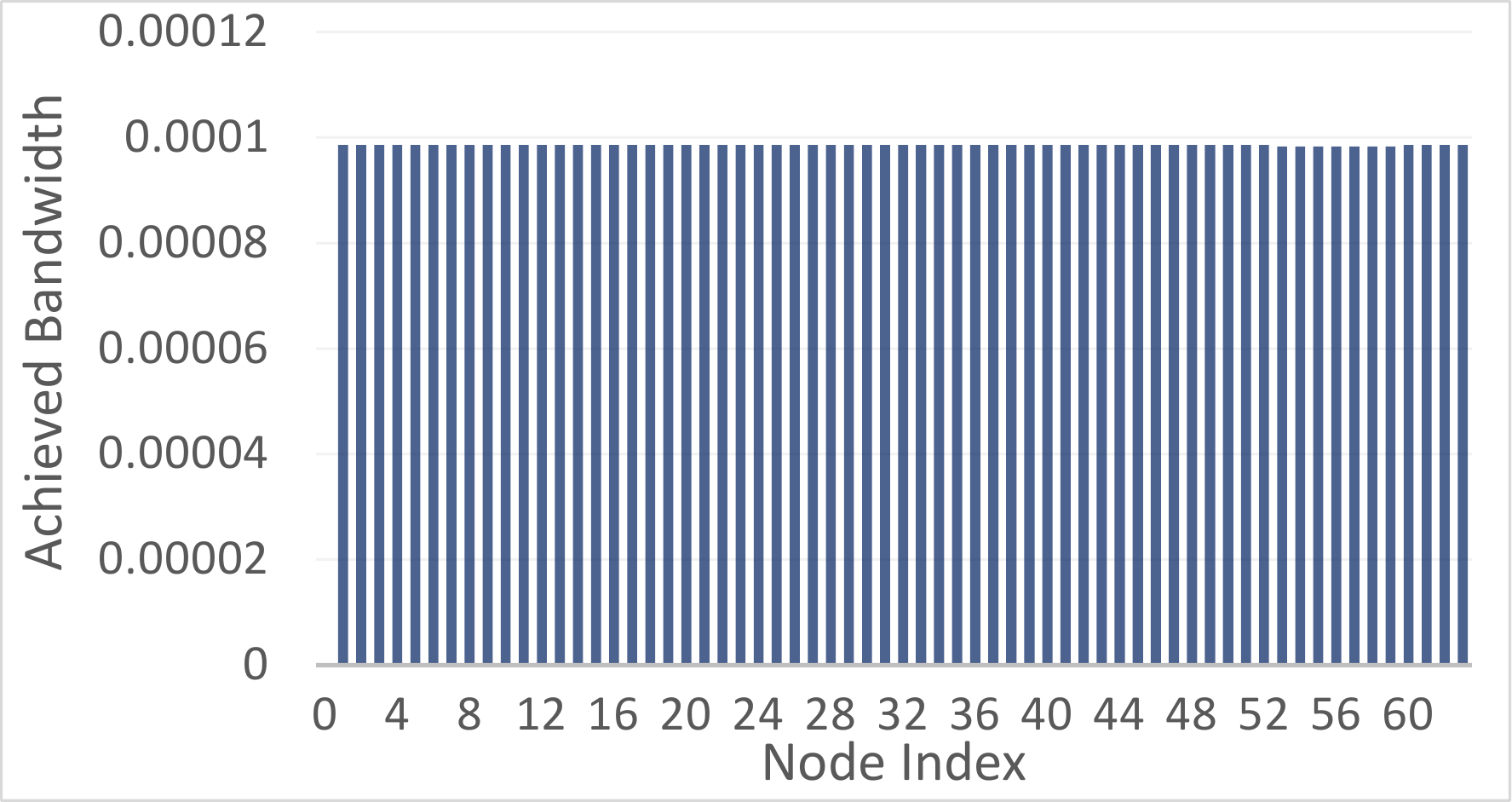}
		\label{fig:FFHotspotInd}
	}
	\caption{Nodal throughput of the nodes in MWMR-AC and Corona}
	\label{fig:indThroughput}
\end{figure}

Figure \ref{fig:indThroughput} illustrates the corresponding results. As shown in Figure \ref{fig:ACHotspotIndDiffW}, the nodal throughput seen by each node is proportional to its weight parameter. The bandwidth achieved by the nodes in the second and third categories is respectively two and three times higher than that of the nodes in the first category. Moreover, nodes belonging to the same category have achieved equal nodal throughput. Furthermore, as shown in Figure \ref{fig:ACHotspotIndSameW}, the nodal throughput seen by all nodes are the same when all categories have the same value for the weight.

For Corona/Corona-FF, we only have the results with all nodes having the same weight since Corona/Corona-FF does not provide any weighting or service differentiation capability. In fact, our admission control policy has the advantage of enabling service differentiation among nodes as shown in Figure \ref{fig:ACHotspotIndDiffW}. Figure \ref{fig:noFFHotspotInd} shows that Corona without Fast Forward token mechanism has an unfair nodal throughput across nodes. Nodes closer to node 0 achieve higher bandwidth than the farther ones. Specifically, nodes 61, 62, and 63 are totally starved. Corona-FF however, does provide a fair nodal throughput.
\subsection{Parsec and SPLASH-2 Simulation Results}
For the benchmark simulations, we use Sniper \cite{carlson2011sniper}, an x86 multicore simulator, to generate PARSEC and SPLASH-2 traffic traces. The traces are then injected into our simulator to obtain the results. In Sniper, we consider 64 nodes, each having its own private L1/L2 (32kB/512kB) and shared L3 caches (1MB). We also consider a 1 (resp.~3) cycle latency to access the tag (resp.~data) of L1 cache (4-way), a 3 (resp.~9) cycle latency to access the tag (resp.~data) of L2 cache (8-way), a 4 (resp.~11) cycle latency to access the tag (resp.~data) of L3 cache (16-way), a cache line size of 64 bytes, and a 100 $\mathrm{ns}$ latency to access the main memory. Furthermore, 8 memory controllers are used for accessing the main memory.
\begin{figure}
	\begin{center}
		\includegraphics[angle=0, scale=0.7]{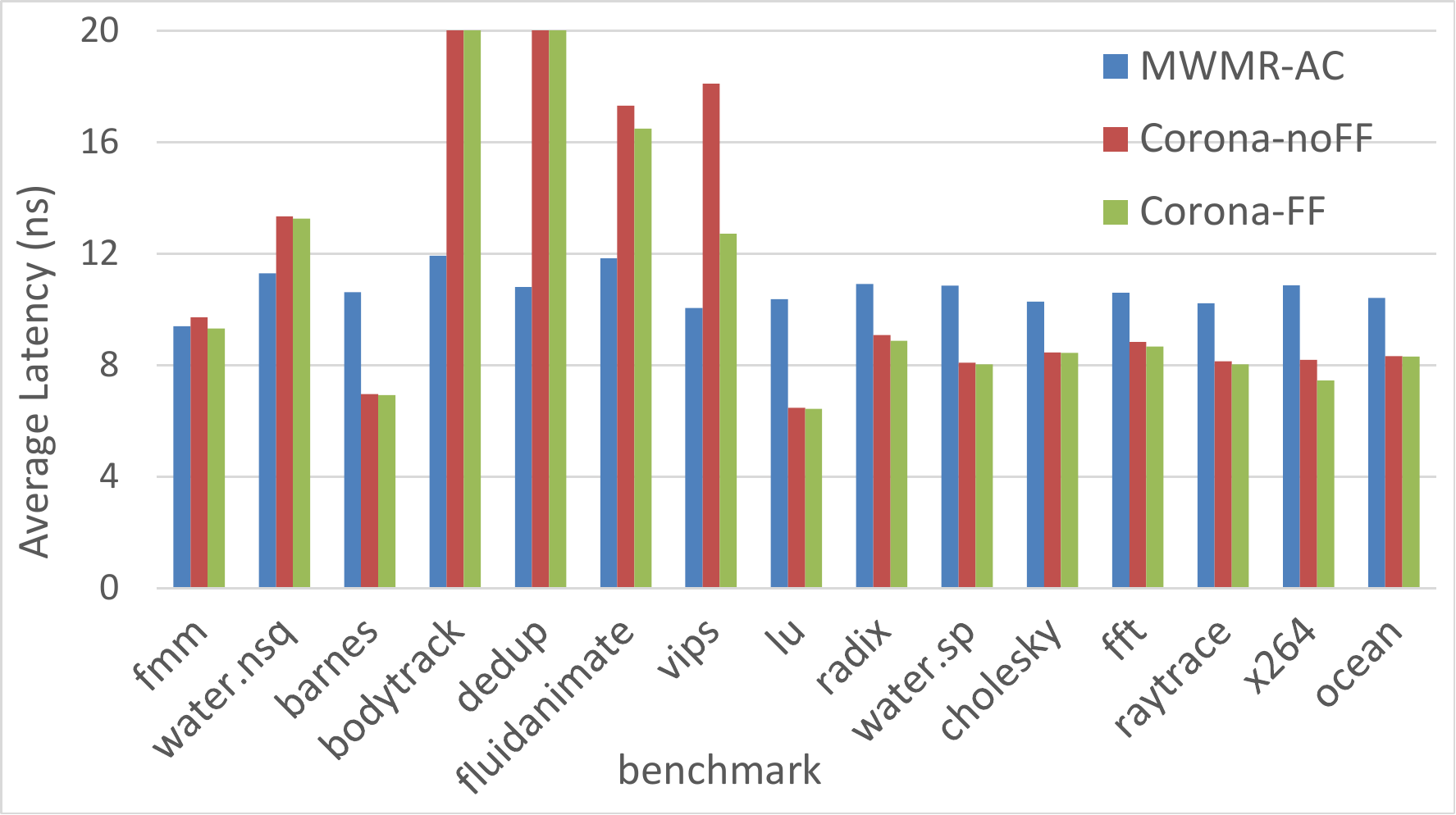}
	\end{center}
	\caption{Average latecy of packets for various benchmarks from SPLASH-2 and PARSEC benchmark suites.}
	\label{fig:SPLASH-PARSEC}
\end{figure}

Figure \ref{fig:SPLASH-PARSEC} shows the total average latency of packets for each benchmark. It can be seen that MWMR-AC outperforms Corona and Corona-FF in 5 of the benchmarks, i.e., water.nsq, bodytrack, dedup, fluidanimate, and vips. In particular, for bodytrack and dedup, we see a significant decrease (4.7x and 6.04x) in latency compared to Corona and Corona-FF. For the rest of the benchmarks, both Corona and Corona-FF provide lower latency than MWMR-AC. This is mainly because these benchmarks impose lower traffic loads on the network in which case the latency of the controller in MWMR-AC becomes the bottleneck. This also explains why we see very close latency results (about 10 to 11 ns) across these group of benchmarks for MWMR-AC. Note that essentially, a non-trivial admission control policy cannot provide much of a benefit compared to a trivial resource allocation mechanism when the traffic load is low. This is because with a low traffic, there is no contention for resources. In such cases, a fast and non-iterative solution such as the one provided in Algorithm \ref{alg:burst_solution} is more desirable. Although we use Algorithm \ref{alg:burst_solution} for low traffic loads, we have considered the overhead of this algorithm equal to one iteration in Algorithm \ref{alg:iterative_solution}. We believe that in practice, such overhead could be considerably lower for Algorithm \ref{alg:burst_solution}, which will in turn improve the performance of MWMR-AC for low-traffic benchmarks.

\section{Conclusion}
\label{sec:conclusion}
Usage of WDM techniques in an optical crossbar provides a huge pool of wavelengths to be shared among competing on-chip cores. In order to manage such resources in a fair and efficient manner, we presented an admission control scenario in an optical on-chip Multiple Write, Multiple Read crossbar. In order to take into account the perceived satisfaction of cores when transmitting data, we cast the problem of wavelength assignment and buffer management as a utility-based convex optimization problem, also referred to as admission control problem. This formulation allowed us to devise a fair and efficient wavelength assignment and buffer management algorithm as the solution to admission control optimization problem. Running on a central admission controller, the proposed algorithm not only tries to achieve the maximum utilization of data channels, but also provides a mechanism to control the access of on-chip nodes to the shared communication resources. 

\bibliographystyle{elsarticle-num}
\bibliography{ONoC-DB}

\end{document}